\documentclass[aps,pre,preprint,onecolumn,citeautoscript,superscriptaddress,floatfix]{revtex4-1}  
\usepackage[utf8]{inputenc}
\usepackage{amsmath,amssymb,mathrsfs,bm,setspace,xspace,soul}
\usepackage{graphicx}  
\usepackage{color} 
\usepackage[papersize={8.5in,11in}]{geometry}

\usepackage{multirow}
\usepackage{amsfonts}
\usepackage{mathtools} 
\usepackage{feynmp}
\usepackage{braket}
\usepackage{tabu}
\usepackage{dsfont} 

\newcommand{\eq}[1]{\begin{align}#1\end{align}}
\newcommand{\eql}[1]{\begin{align}#1\end{align}}
\newcommand{\tnl}{\\[1.5ex]}
\newcommand{\ttnl}{\\[1.2ex]}
\usepackage[colorlinks=true]{hyperref} 
\hypersetup{
    bookmarks=true,         
    unicode=false,          
    pdftoolbar=true,        
    pdfmenubar=true,        
    pdffitwindow=false,     
    pdfstartview={FitH},    
    pdftitle={My title},    
    pdfauthor={Author},     
    pdfsubject={Subject},   
    pdfcreator={Creator},   
    pdfproducer={Producer}, 
    pdfkeywords={keyword1} {key2} {key3}, 
    pdfnewwindow=true,      
    colorlinks=true,       
    linkcolor=magenta, 
    citecolor=blue,        
    filecolor=magenta,      
    urlcolor=cyan           
} 

\newcommand{\beq}{\begin{equation}}
\newcommand{\eeq}{\end{equation}}

\newcommand{\bQ}{{\bm Q}}

\def\bea{\begin{eqnarray}}
\def\eea{\end{eqnarray}}

\begin{document}

\newcommand{\ord}{\mathcal{O}}
\newcommand{\ham}{\hat{H}}
\newcommand{\hamk}{\mathcal{H}}
\newcommand{\hilb}{\mathscr{H}}
\newcommand{\M}{\mathcal{M}}
\newcommand{\vn}{{\mathbf{n}}}
\newcommand{\vk}{{\mathbf{k}}}
\newcommand{\vQ}{{\mathbf{Q}}}
\newcommand{\vK}{{\mathbf{K}}}
\renewcommand{\vr}{{\mathbf{r}}}
\newcommand{\vrr}{{\mathbf{r}}}
\newcommand{\vq}{{\mathbf{q}}}
\newcommand{\vp}{{\mathbf{p}}}
\newcommand{\vv}{{\mathbf{v}}}
\newcommand{\hx}{\hat{\mathbf{x}}}
\newcommand{\hy}{\hat{\mathbf{y}}}
\newcommand{\J}{\mathcal{J}}
\newcommand{\V}{\mathcal{V}}
\renewcommand{\P}{\mathcal{P}}
\newcommand{\T}{\mathcal{T}}
\newcommand{\vphi}{{\varphi}}
\newcommand{\bchi}{\hat{\bb{\chi}}}
\newcommand{\tphi}{\tilde{\phi}}
\newcommand{\bcX}{\boldsymbol{\mathcal{X}}}
\newcommand{\bcW}{\boldsymbol{\mathcal{W}}}

\let\spt\mathcal
\let\v\mathbf
\newcommand{\ep}{\epsilon}
\renewcommand{\a}{\alpha}
\renewcommand{\b}{{\beta}}
\renewcommand{\d}{\delta}
\newcommand{\g}{\gamma}
\newcommand{\n}{\nu}
\newcommand{\m}{\mu}
\renewcommand{\t}{\tau}
\newcommand{\s}{\sigma}
\newcommand{\w}{\omega}
\renewcommand{\(}{\left(}
\renewcommand{\)}{\right)}
\renewcommand{\[}{\left[}
\renewcommand{\]}{\right]}
\renewcommand{\dag}{\dagger}
\newcommand{\ninfty}{{-\infty}}
\newcommand{\ua}{\uparrow}
\newcommand{\da}{\downarrow}
\newcommand{\nt}{\notag}
\newcommand{\ph}{\phantom}
\newcommand{\id}{\mathds{1}}
\newcommand{\ti}{\tilde}
\newcommand{\bb}{\boldsymbol}
\renewcommand{\o}{\over}

\title{Charge ordering in three-band models of the cuprates}  
 \author{Alexandra Thomson}
 \affiliation{Department of Physics, Harvard University, Cambridge, Massachusetts, 02138, USA}
 \author{Subir Sachdev}
 \affiliation{Department of Physics, Harvard University, Cambridge, Massachusetts, 02138, USA}
 \affiliation{Perimeter Institute for Theoretical Physics, Waterloo, Ontario N2L 2Y5, Canada}
 \date{\today\\
 \vspace{0.6in}}
\begin{abstract} 
We examine trends in the wavevectors and form-factors of charge density wave instabilities of three-band models
of the underdoped cuprates. For instabilities from a high temperature state with a large Fermi surface, we 
extend a study by Bulut {\em et al.} (Phys. Rev. B {\bf 88}, 155132 (2013)) to include a direct antiferromagnetic
exchange coupling between the Cu sites. As in previous work, we invariably find
that the primary instability has a diagonal wavevector $(\pm Q_0, \pm Q_0)$ and a $d$-form factor.
The experimentally observed wavevectors along the principal axes $(\pm Q_0,0)$, $(0, \pm Q_0)$ have higher energy, but they also have a predominantly $d$-form factor.  
Next, we gap out the Fermi surface in the anti-nodal regions of the Brillouin zone by including static, long-range antiferromagnetic 
order at the wavevector $(\pi, \pi)$: this is a simple model of the pseudogap in which we assume the antiferromagnetic order averages to zero by
`renormalized classical' thermal fluctuations in its orientation, valid when the antiferromagnetic correlation length is large.
The charge density wave instabilities of this pseudogap state 
are found to have the optimal wavevector $(\pm Q_0,0)$, $(0, \pm Q_0)$, with the magnitude of the $d$-form factor decreasing with increasing magnetic order. 
\end{abstract}
\maketitle 
\tableofcontents 
\section{Introduction}
\label{sec:intro}

A number of recent scanning tunneling microscopy (STM) and X-ray scattering 
experiments have provided interesting new information on the microstructure of the charge order at wavevectors $(\pm Q_0,0)$, $(0, \pm Q_0)$
in the hole-doped cuprates (here $Q_0$ ranges between $2\pi/3$ and $2\pi/4$). 
The STM observations by Fujita {\em et al.\/} \cite{Fujita2014} on Bi$_2$Sr$_2$CaCu$_2$O$_{8+x}$ 
and Ca$_{2-x}$Na$_x$CuO$_2$Cl$_2$ yield direct phase-sensitive evidence of a dominant
$d$-form factor density wave. Comin {\em et al.\/} \cite{Comin14} performed X-ray scattering off the Cu sites in YBa$_2$Cu$_3$O$_{6+y}$;
interpretation of their results require a model of the density wave distribution around the Cu sites, and this model yields the best fit  
with a significant $d$-form factor. In contrast, in the La-based superconductor La$_{1.875}$Ba$_{0.125}$CuO$_4$,
Achkar {\em et al.\/} \cite{Achkar14} performed X-ray scattering off the O sites, and their results are directly interpreted in terms of a 
dominant $s'$ form factor. In this context, it will be important for our purposes to note that the La-based superconductors,
with the $s'$ form factor, have long-range incommensurate magnetic order at low temperatures, while the other superconductors do not.

On the theoretical side, a number of recent studies have investigated density wave instabilities with form factors carrying non-zero angular momentum \cite{MMSS10,metzner1,metzner2,yamase,SSRP13,kee,SSJS14,DHL13,HMKE,HMKE13,HF14,SWSS14,AASS14,AASS14b,DCSS14,YWAC14,norman14,ATAC14,akb13,3band,EAK14}.
It is important to note that in our discussion form factors are defined using the expression
\beq
\left \langle c_{i \alpha}^\dagger \, c_{j \alpha}^{\vphantom \dagger} \right\rangle = 
\sum_{\vQ} \left[ \sum_{\vk} P_{\vQ} ( \vk) e^{i \vk \cdot (\vrr_i - \vrr_j)} \right] e^{i \vQ \cdot (\vrr_i + \vrr_j)/2},
\label{eqn:oneband}
\eeq
or
\beq 
\left \langle c_{\vk + \vQ/2}^\dagger \, c_{\vk - \vQ/2}^{\vphantom \dagger} \right\rangle = P_{\vQ} (\vk) ,
\eeq
for the case of a single-band model (with generalizations to multi-band models to be discussed below); 
here $c_{i \alpha}$ annihilates an electron with spin $\alpha$ on the Cu site $i$, $\vQ$ is the ordering wavevector, 
and $P_\vQ (\vk)$ is the form factor. The form factor is required to obey $P_{-\vQ} (\vk) = P_\vQ^\ast (\vk)$, 
while time-reversal symmetry imposes $P_\vQ (-\vk) = P_\vQ (\vk)$. In computations starting from a Fermi liquid with a large Fermi surface
in a single band model, it was found that the dominant density wave instability was at wavevectors $\vQ = (\pm Q_0, \pm Q_0)$ with a $d$-form
factor $P_\bQ (\vk) \sim \cos (k_x) - \cos (k_y)$. The ordering wavevector of these instabilities is therefore along the diagonal of the square
lattice Brillouin zone, rather than the along the principal axes as observed in the experiments.
An extension of these computations to the 3-band model by Bulut {\em et al.\/} \cite{akb13} and by 
Maier and Scalapino \cite{Maier} also found the diagonal wavevector. However, these 3-band computations did not include a direct 
antiferromagnetic exchange interactions between the Cu orbitals; such an exchange was crucial in the arguments for the $d$-form factor using
the pseudospin rotation symmetry to the $d$-wave superconductor. The present paper will extend the 3-band computations to include
 Cu-Cu exchange interactions: this significantly increases the computational complexity because the particle-hole diagrams have 
off-site interactions. The results for such computations appear in Section~\ref{sec:lfs}: we find that the ordering wavevectors remain along the diagonals,
as in the previous 3-band computations. However, we do obtain new information on the off-site correlators characterizing the density wave, and all are
found to be in excellent accord with a $d$-form factor interpretation.

A number of proposals have been made to resolve the disagreement between theory and experiment in the orientation of the
wavevector \cite{AASS14,DCSS14,HMKE13,3band}. 
In particular, Atkinson {\em et al.} \cite{3band} have argued that it is important to examine the charge ordering instabilities of a Fermi surface
with pre-existing `pseudogap', and not of the large Fermi surface. They proposed to induce an analog of the pseudogap by imposing
commensurate antiferromagnetic order at the wavevector $(\pi, \pi)$ on the parent state; from this parent state they
found that the optimal charge-ordering wavevector was indeed similar to the experimentally observed values of $(\pm Q_0,0)$, $(0, \pm Q_0)$ along the principal axes. In reality, there is no antiferromagnetic order in the parent state of the hole-doped superconductors, 
but such a `renormalized classical'
approach may be justified if the antiferromagnetic correlation length is large enough \cite{Tremblay04}. We will also take such a model of the pseudogap in the present paper, extended to our 3-band model with a bare Cu-Cu exchange interaction. 
Our analysis, presented in Section~\ref{sec:sfs}, will also allow for the mixing present between the charge order at $\vQ$ and 
spin density wave order at $\vQ + (\pi, \pi)$, and diagonalize the eigenmodes in the full charge-spin space. 
Our computations also find that the optimal charge ordering wavevector is close to the experimentally observed values of $(\pm Q_0,0)$, $(0, \pm Q_0)$ along the principal axes. Another finding is that the presence of antiferromagnetic order decreases the 
magnitude of the $d$-form factor; this trend is consistent with recent observation of a dominant $s'$ form factor in the hole-doped cuprate with magnetic order, La$_{1.875}$Ba$_{0.125}$CuO$_4$ \cite{Achkar14}.

A weakness of the above antiferromagnetic model of the pseudogap is, of course, that the antiferromagnetic correlation length is actually quite short in the hole-doped cuprates. This suggests that one should include quantum spin fluctuations more fully, and account better
for `spin liquid' physics. The computation described above can be regarded as one limiting case where the spin fluctuations are presumed to be fully thermal and classical. The opposite limiting case is one where the spin fluctuations are fully quantum, and the pseudogap  is due to a spin liquid background: such a perspective was taken in a separate paper \cite{DCSS14}, which finds a predominant $d$-form factor 
and an optimal wavevector of $(\pm Q_0,0)$, $(0, \pm Q_0)$ along the principal axes, both in agreement with experiments.

\section{Large Fermi surface}
\label{sec:lfs}

This section will examine the density wave instabilities of the 3-band model of the CuO$_2$ layers of the cuprates. Here we will start
from a Fermi liquid ground state {\em without\/} any magnetic order.

\begin{table}
\begin{minipage}{\textwidth}\center
\begin{tabular}{l |*{4}{c} | *{5}{c}}
\centering
 \multirow{2}{*}{Parameters} & \multicolumn{4}{c|}{Hopping} & \multicolumn{5}{c}{Interactions}\\ \cline{2-10}
 & $t_{pd}$ & $t^d_{pp}$ & $t^i_{pp}$ & $\ep_d-\ep_p$ & $U_d$ & $U_p$ & $V_{pd}$ & $V_{pp}$ & $J$ \\ \hline\hline
Large Fermi Surface & 1.6 & 0.0 & -1.0 & 0.9 & 9.0 & 3.0 & 1.0 & variable &variable \\
Small Fermi Surface & 1.6 & 0.0 & -1.0 & 0.9 & variable & 0.0 & 1.0 & 1.5 & variable 
\end{tabular}\end{minipage}
\caption{Parameters used in the calculations presented in Sections \ref{sec:lfs} and \ref{sec:sfs} in units where $|t_{pp}^i|=1$. With the exception of $J$, the parameters are the same as those given in Ref.~\onlinecite{3band} for the large Fermi surface calculation. For the small Fermi surface, $U_d$ and $J$ were determined by a Hartree-Fock analysis.}\label{tab:param}
\end{table}

\begin{figure}
\centering
\includegraphics[scale=0.7]{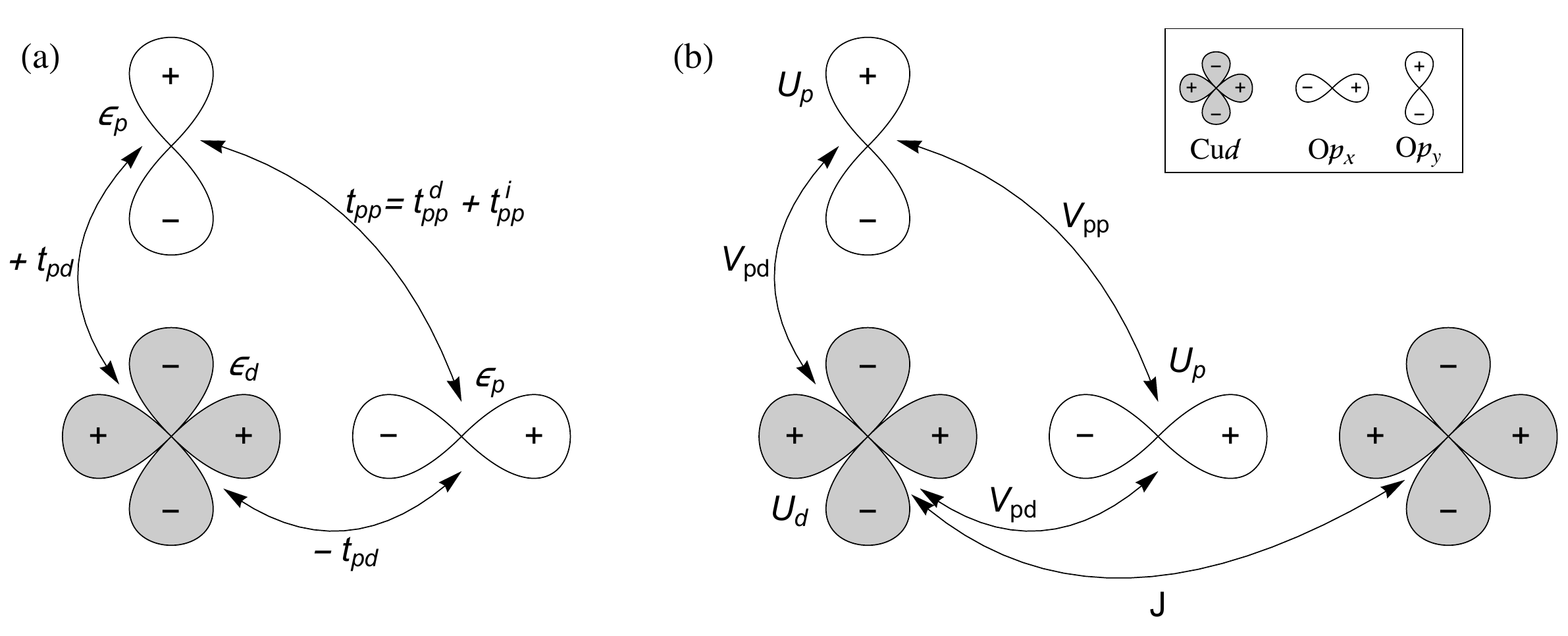}
\caption{(a) Diagram of the unit cell of the lattice showing the hopping amplitudes corresponding to the tight-binding Hamiltonian in Eq.~\ref{eqn:kernel}. The on-site energies $\ep_d$ and $\ep_p$ and the inter-orbital hopping parameters $t_{pd}$ and $t_{pp}$ are displayed next to the corresponding orbitals and bonds respectively. Note how the sign of $t_{pd}$ changes depending on the relative phases of the wavefunction lobes closest to one another. (b) Diagram showing the interactions of the Hamiltonian within the unit cell and between two nearest-neighbor copper atoms. The corresponding expressions are given in Eqs.~\ref{eqn:coulomb} and \ref{eqn:exchange}. }
\label{fig:hoppingintparams}
\end{figure}

Following Ref.~\onlinecite{3band}, the hopping Hamiltonian $\ham_t$ we we use to describe this ground state is given by an extension of the Emery model \cite{emery} due to Andersen \emph{et al.} \cite{ALJP}: 
\eq{
\ham_t&=\sum_\vk \Psi^\dagger_{\vk,\a}\hamk(\vk)\Psi_{\vk,\a},\qquad\qquad \Psi_{\vk,\a}^\dag=\left(c_{d\a}^\dag(\vk),c_{x\a}^\dag(\vk),c_{y\a}^\dag(\vk)\right)\notag\\
\hamk(\vk)&=
\begin{pmatrix}
\ep_d & 2t_{pd}\sin(k_x/2) & -2t_{pd}\sin(k_y/2)\\
2t_{pd}\sin(k_x/2) & \ep_p+4t^i_{pp}\sin^2(k_x/2) & 4(t^d_{pp}+t^i_{pp})\sin(k_x/2)\sin(k_y/2) \\
-2t_{pd}\sin(k_y/2) & 4(t^d_{pp}+t^i_{pp})\sin(k_x/2)\sin(k_y/2)  & \ep_p+4t^i_{pp}\sin^2(k_y/2) 
\end{pmatrix}\,.
\label{eqn:kernel}}
For $t_{pp}^i=0$, this Hamiltonian reduces to the Emery model and corresponds to the tight-binding model of the unit cell shown in Fig.~\ref{fig:hoppingintparams}(a). The signs of the Cu-O hopping amplitudes are determined by the phases of neighboring lobes of the orbital wavefunctions. 
In order to obtain a realistic Fermi surface for the cuprates, however, we must impose that the direct hopping amplitude between oxygen orbitals $t_{pp}^d$ be negative by hand. This is unsatisfactory since we would normally expect that $t_{pp}^d$ be positive. 
By integrating out the Cu 4$s$ orbital from a 4-band Hamiltonian, Andersen \emph{et al.} found a negative indirect hopping amplitude $t_{pp}^i$. Further, the direct hopping amplitude they calculated was comparatively small. 
Consequently, in what follows, we have set $t_{pp}^i=-1.0$ and $t_{pp}^d=0$. The remaining values of the hopping amplitudes are given in Table~\ref{tab:param}. We note that for computational ease we have chosen our gauge, given in Eq.~(\ref{eqn:fourdef}) in Appendix \ref{app:timerev}, so that $\ham_t$ is a real symmetric matrix. 

The Green's functions are given by diagonalizing the kernel $\hamk(\vk)$:
\eq{
\mathbf{S}^\dag(\vk)\hamk(\vk)\mathbf{S}(\vk)=\mathbf{\Lambda}(\vk)
}
where $\Lambda_{\mu\nu}(\vk)=\d_{\mu\nu}E^\mu_\vk$ gives the band energies and $\mathbf{S}(\vk)$ is a $3\times3$ matrix of eigenvectors. 
In the diagonal basis, the bare Green's function is 
\eq{
\mathcal{G}^0_{\g}(\vk;\omega_n)={-1\o i\omega_n-(E^\g_\vk-\mu)}
}
and so the Green's function in the orbital basis is
\eq{
G^0_{\mu\nu}(\vk;\omega_n)
&=-\sum_\g S^*_{\mu\g}(\vk)S_{\nu\g}(\vk){1\o i\omega_n-(E^\g_\vk-\mu)}\,.
}

We consider the effect of two types of interactions. The first is the Coulomb interaction $\ham_C$, which we further separate into a Hubbard term, $\ham_C^h$, and an inter-orbital term, $\ham_C^v$:
\eq{\label{eqn:coulomb}
\ham_C=& \ham_C^h+\ham_C^v \\ 
\ham_C^h	=&\sum_i \nt
\left[ U_d \,c_{d\ua}^\dag(\vr_i)c_{d\ua}(\vr_i)c_{d\da}^\dag(\vr_i)c_{d\da}(\vr_i) +
U_p \left( c_{x\ua}^\dag(\vr_i)c_{x\ua}(\vr_i)c_{x\da}^\dag(\vr_i)c_{x\da}(\vr_i) +
                c_{y\ua}^\dag(\vr_i)c_{y\ua}(\vr_i)c_{y\da}^\dag(\vr_i)c_{y\da}(\vr_i) \right)
                \right] \\ \nt
\ham_C^v=&\sum_{\braket{ij}} V_{pd}c_{d\a}^\dag(\vr_i)c_{d\a}(\vr_i)
	\left[ c_{x\b}^\dag(\vr_j)c_{x\b}(\vr_j)  + c_{y\b}^\dag(\vr_j)c_{y\b}(\vr_j)  \right] \\ \notag
&+\sum_{\braket{ij}} V_{pp}c_{x\a}^\dag(\vr_i)c_{x\a}(\vr_i)c_{y\b}^\dag(\vr_j)c_{y\b}(\vr_j) \notag}
where the sums in the last two lines are over nearest-neighbors. We go beyond the previous work \cite{3band,akb13} by also including
a direct exchange term between the Cu atoms
\eq{\label{eqn:exchange}
\ham_J=\sum_{\braket{ij}}\sum_a{J\o4}\s^a_{\a\b}\s^a_{\gamma\delta}c_{d\a}^\dag(\vr_i) c_{d\b}(\vr_i)c_{d\gamma}^\dag(\vr_j) c_{d\delta}(\vr_j) \,,
}
where the sum is over nearest-neighbor interactions between Cu atoms in different unit cells. The interactions between the various orbitals are shown in Fig.~\ref{fig:hoppingintparams}(b). The full Hamiltonian is given by the sum 
\eq{\label{eqn:ham}
\ham=\ham_t+\ham_C+\ham_J\,.
}

We transform the interaction terms to momentum space, and express them using a suitable set of basis functions in 
Appendix~\ref{app:timerev}. 
For the particle-hole singlet channel which we wish to study, 23 basis functions are required. This can easily be seen by counting the number on-site and inter-orbital interactions present in the Hamiltonian. The Hubbard term has on-site copper, O $p_x$, and O $p_y$ orbital interactions and, since they are local, they require one basis function each. Conversely, the inter-orbital interactions have two separate degrees of freedom. For instance, each copper atom per unit cell interacts through the Coulomb term with 4 distinct oxygen orbitals, and so requires 8 basis functions. Similarly, the interactions among the O $p_x$ and O $p_y$ orbitals and between the copper orbitals in different unit cells introduce another 12 basis functions. These are given in Table \ref{tab:basis} in Appendix \ref{app:timerev}.

As mentioned above, we are primarily interested in the density wave instability of this model in the particle-hole channel. In order to do study this, we generalize the order parameter defined in Eq.~(\ref{eqn:oneband}) to the 3-band model by the addition of orbital indices.
Accounting for the gauge choice given in Eq.~(\ref{eqn:fourdef}), we write
\eq{\label{eqn:orderparam}
P_{ij}^{\m\n}&=\Braket{c_{\mu\a}^\dag(\vr_i)c_{\nu\a}(\vr_j)} ={\mathfrak{z}_{\m\n}}\sum_\vQ\left[\int {d^2k\o4\pi^2}\, P^{\mu\nu}_\vQ(\vk)e^{i\vk\cdot(\vr_i-\vr_j)}e^{i\vk\cdot(\v{R}_\mu-\v{R}_\nu)}\right]e^{i\vQ\cdot(\vr_i+\vr_j)/2}e^{i\vQ\cdot(\v{R}_\mu+\v{R}_\nu)/2} \\
 \mathfrak{z}_{\m\n}&=\left\{\begin{matrix*}[l]
\ph{-}1,	&  	\mu\nu=dd,xx,yy,xy,yx\\
-i,	&	\m\n=dx,dy \\
\ph{-}i,	&	\m\n=xd,yd 
\end{matrix*}\right. \nt\,.}
where there is no implied summation of $\m$ and $\n$. In momentum space, the order parameter $P^{\mu\nu}_\vQ(\vk)$ can also be decomposed into the basis functions given in Table \ref{tab:basis}:
\eq{\label{eqn:basisexpansion}
P^{\m\n}_\vQ(\vk)&= \mathfrak{z}_{\m\n}\sum_{l=1}^{23}\P_l(\vQ)\phi^{\,l}_{\m\n}(\vk)\,. 
}
Hermiticity requires that $P^{\mu\nu}_{ij}=\left(P^{\nu\mu}_{ji}\right)^*$ and it follows that in momentum space
\eq{\label{eqn:hermiticity}
P^{\m\n}_{\vQ}(\vk)=\(P^{\n\m}_{-\vQ}(\vk)\)^* .
}
Because of the Fourier definition in Eq.~(\ref{eqn:fourdef}), time reversal $\T$ acts on the electron annihilation operator $c_{\mu\a}(\vk)$ as 
\eq{
\T c_{\mu\ua}(\vk) \T^{-1} = \eta_{\mu\nu}c_{\mu\da}(-\vk),\qquad\qquad \eta=\text{diag}(1,-1,-1). 
} 
It follows that the order parameter transforms as
\eq{\label{eqn:timerev}
\T: P^{\m\n}_{\vQ}(\vk)\mapsto \eta_{\mu\g}\eta_{\n\d}P^{\g\d}_{\vQ}(-\vk) \,.
}
The action of $\T$ on the functions $\P_l(\vQ)$ is summarized in Table \ref{tab:timerev} of Appendix \ref{app:timerev}.
\subsection{Particle-hole interactions}\label{sec:Tmatrix}

This subsection will compute the particle-hole ladder diagrams associated with density wave instabilities, and find their eigenmodes (the components $\{\P_l(\vQ)\}$ defined in Eq.~\ref{eqn:basisexpansion}) as a function
of the total momentum of the particle-hole pair.

\begin{figure}
\includegraphics[scale=1.0]{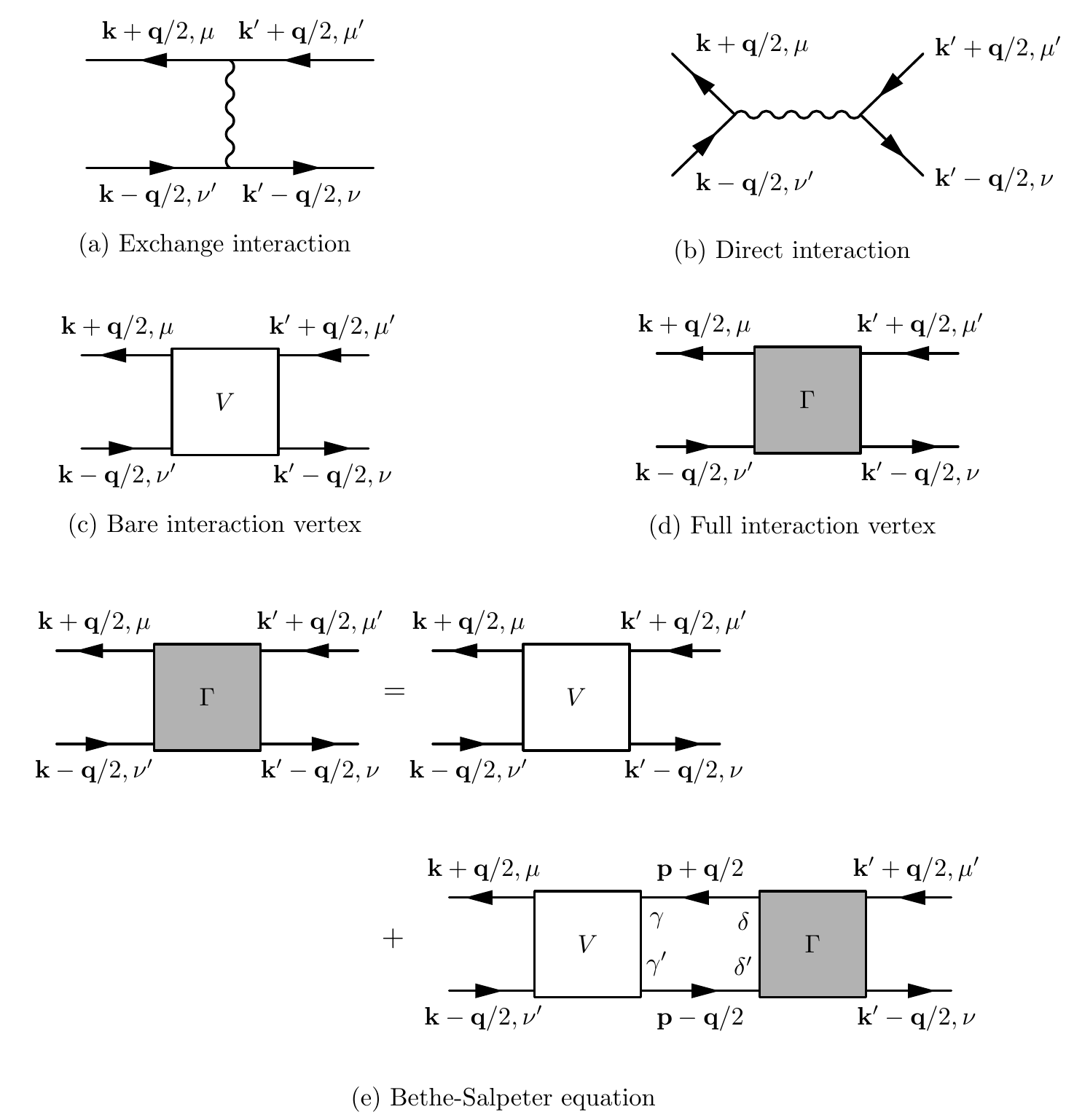}
\caption{Feynman diagrams used in the $T$-matrix calculation discussed in Section \ref{sec:Tmatrix}. (a) and (b) give the exchange and direct interactions respectively. Together, they compose the bare vertex shown in (c) as per Eq.~\ref{eqn:barevertex}. (d) shows the full interaction vertex, which is determined by the Bethe-Salpeter equation given diagrammatically in (e). Further details are presented in the main text.}\label{feynall}
\end{figure}

Similar to Ref. \cite{akb13}, we accomplish this by reducing the Bethe-Salpeter equation to a matrix equation and numerically solving. We start by defining an effective interaction as a sum between the exchange and direct interactions for the charge channel:
\eq{
\mathbf{V}_{\mu\mu',\n\n'}(\vk,\vk',\vq)=\mathbf{X}_{\mu\mu',\n\n'}(\vk-\vk')-2\mathbf{W}_{\mu\mu',\n\n'}(\vq)\,.
\label{eqn:barevertex}}
The exchange $\mathbf{X}_{\mu\mu',\n\n'}$ and direct $\mathbf{W}_{\mu\mu',\n\n'}$ parts of the interaction are represented in (a) and (b) of Fig.~\ref{feynall} respectively and the bare interaction $\mathbf{V}_{\mu\mu',\n\n'}$ corresponds to Fig.~\ref{feynall}(c).
In terms of the 23 basis functions, we write the exchange vertex as
\eq{
\mathbf{X}_{\mu\mu',\n\nu'}(\vk-\vk')=\sum_{l,m=1}^{23}\phi_{\mu\nu'}^{\,l}(\vk)X_{lm}\phi_{\mu'\nu}^{\,m}(\vk')}
where $X_{lm}=\V_l\d_{lm}$ and $\V_l$ are interaction parameters whose relation to the values presented in Table~\ref{tab:param} are given in Eq.~(\ref{eqn:intparam}). 
Similarly, the direct vertex is expressed as
\eq{
\mathbf{W}_{\mu\mu',\n\n'}(\vq)=\sum_{l,m=1}^{23}{\phi_{\mu\nu'}^{l}}\,W_{lm}(\vq)\phi_{\mu'\nu}^{\,m}
}
where for $l,m>3$, $W_{lm}(\vq)=0$ and for $l,m=1,2,3$, it is given by
{\renewcommand*{\arraystretch}{1.3}\eq{\label{eqn:directint}
W_{lm}(\vq)=
\begin{pmatrix}
U_d & 2V_{pd}\cos(q_x/2) & 2V_{pd}\cos(q_y/2) \\
2V_{pd}\cos(q_x/2) & U_p & 4V_{pp}\cos(q_x/2)\cos(q_y/2) \\
2V_{pd}\cos(q_y/2) & 4V_{pp}\cos(q_x/2)\cos(q_y/2) & U_p
\end{pmatrix}_{lm}
}}
Note that for $l,m=1,2,3$, the basis functions $\phi^{\,l}_{\mu\nu}$ are indeed independent of $\vk$.
 
The leading instability of the total vertex  $\mathbf{\Gamma}_{\mu\mu',\n\n'}(\vk,\vk',\vq)$, shown in the diagram in Fig.~\ref{feynall}(d), defines the order parameter and ordering wavevector we are interested in.
We approximate it by a generalized RPA (Bethe-Salpeter equation) scheme as
\eql{\label{eqn:bethesalpeter}
\mathbf{\Gamma}_{\mu\mu',\n\n'}(\vk,&\vk',\vq)=\sum_{l,m=1}^{23}\phi^{\,l}_{\mu\nu'}(\vk)\Gamma_{lm}(\vq)\phi^{\,m}_{\mu'\nu}(\vk')\\ \notag
&=\sum_{l,m=1}^{23}\phi^l_{\mu\nu'}(\vk)V_{lm}(\vq)\phi^{\,m}_{\mu'\nu}(\vk') \\
&+\sum_{l,m=1}^{23}\sum_{n,s=1}^{23}\sum_{\substack{\g\g' \\ \d\d'}}\sum_{\vp,\omega_n}
\phi^l_{\mu\nu'}(\vk)V_{ln}(\vq)\phi_{\g\g'}^n(\vp)G^0_{\d'\g'}(\vp-\vq/2;\w_n) G^0_{\g\d}(\vp+\vq/2;\w_n)\phi^s_{\d\d'}(\vp)\Gamma_{sm}(\vq)\phi^m_{\mu'\nu}(\vk').\notag
}
The corresponding diagrammatic expression is shown in Fig.~\ref{feynall}(e).
To simplify, we define the polarizability to be
\eq{
\Pi_{ns}(\vq)&=2\sum_{\substack{\g\g' \\ \d\d'}}\sum_{\vp,\omega_n}\phi_{\g\g'}^n(\vp)G^0_{\d'\g'}(\vp-\vq/2)G^0_{\g\d}(\vp+\vq/2)\phi^s_{\d\d'}(\vp)\\
&=-2\sum_{\vp}\sum_{\substack{\g\g' \\ \d\d'}}\sum_{\a\a'}\phi^{\,n}_{\g\g'}(\vp)\phi^{\,s}_{\d\d'}(\vp)\mathcal{M}^{\d'\g'\g\d}_{\a'\a,\vp\vq}{f(E_{\a'}(\vp-\vq/2))-f(E_\a(\vp+\vq/2))\o E_{\a'}(\vp-\vq/2)-E_{\a}(\vp+\vq/2)}\nt
}
where $f$ is the Fermi function and
\eq{
\mathcal{M}^{\d'\g'\g\d}_{\a'\a,\vp\vq}=S^*_{\d'\a'}(\vp-\vq/2)S_{\g'\a'}(\vp-\vq/2)S^*_{\g\a}(\vp+\vq/2)S_{\d\a}(\vp+\vq/2)\,.
}
As indicated, it follows that Eq.~(\ref{eqn:bethesalpeter}) can be reduced to a matrix equation:
\eq{
\Gamma_{lm}(\vq)&=V_{lm}(\vq)+{1\o2}\sum_{n,s=1}^{23}V_{ln}(\vq)\Pi_{ns}(\vq)\Gamma_{sm}(\vq) \\
&=\V_l\d_{lm}-2W_{lm}(\vq)+{1\o2}\sum_{s=1}^{23} \V_l\Pi_{ls}(\vq)\Gamma_{sm}(\vq)-\sum_{n,s=1}^{23}W_{ln}(\vq)\Pi_{ns}(\vq)\Gamma_{sm}(\vq)\,.\nt
}
The instabilities of the total vertex are determined by finding the minimum eigenvalues $\lambda_\vq$ of the matrix 
\eq{\label{eqn:denom}
A_{lm}(\vq)=\d_{lm}-{1\o2}\sum_{n=1}^{23}V_{ln}(\vq)\Pi_{nm}(\vq)\,.
}
for all $\vq$. The ordering wavevector corresponds to the $\vQ_m$ for which $\lambda_{\vQ_m}$ is the global minimum over the entire Brillouin zone and the order parameter is defined by the associated eigenvector $\{\P_l(\vQ_m)\}$ through Eq.~\ref{eqn:basisexpansion}.
\subsection{Results}

\begin{figure}[p!]
\includegraphics[scale=1.0]{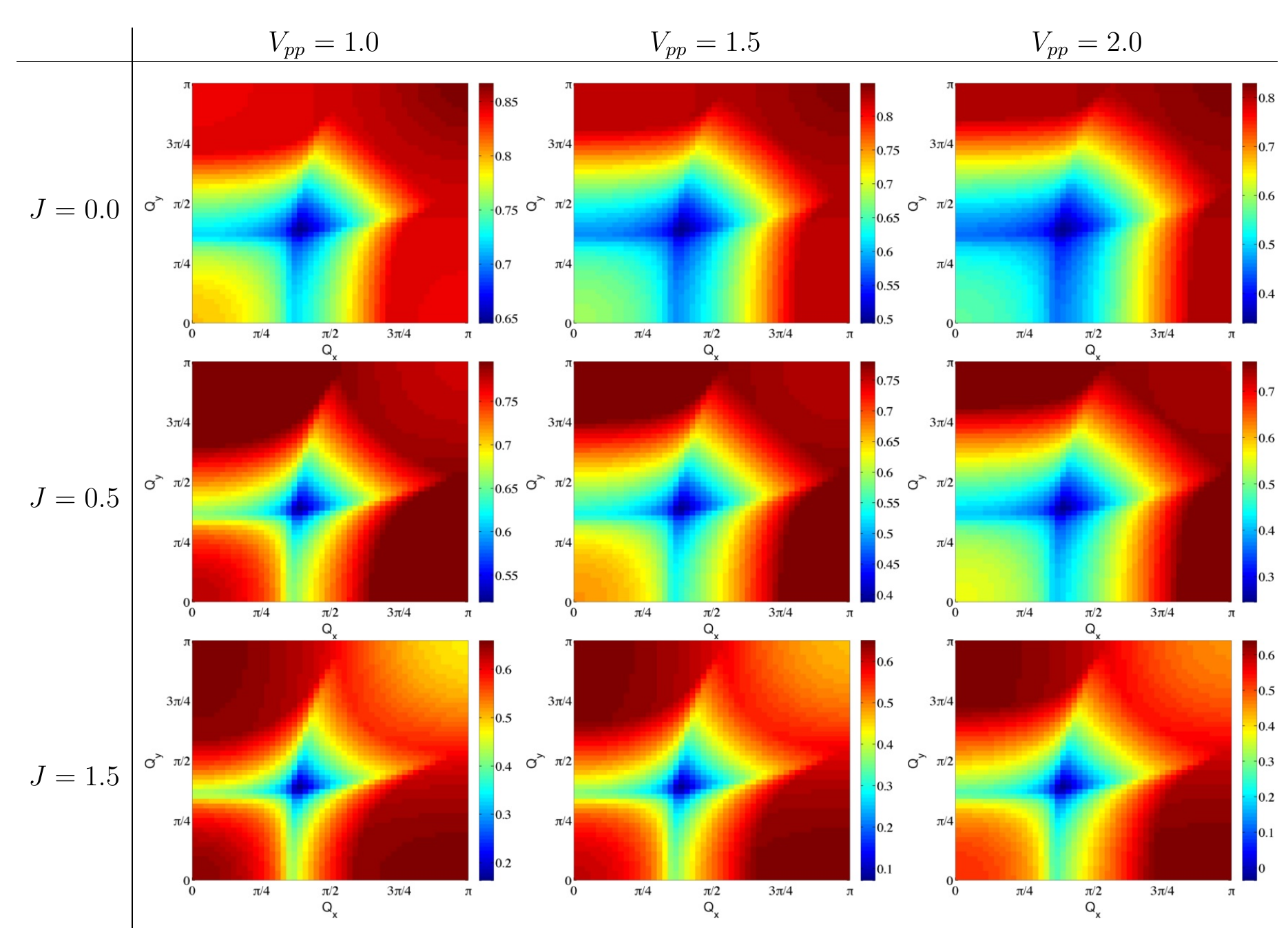}
\caption{(Color online) Plot of the minimum eigenvalue of the matrix $A_{lm}(\vQ)$ in Eq.~(\ref{eqn:denom})  for $\vQ$ in the first quadrant at different values of $J$ and $V_{pp}$. The temperature is $T=0.015$ and the filling $p=5-n=0.1643$. All other parameters are as given in Table \ref{tab:param}. The diagonal point $\vQ=Q_m(1,1)$ for $Q_m=1.19381$ is very consistently the point of greatest instability.}
\label{fig:mvals}
\end{figure}

The lowest eigenvalues of the matrix in Eq.~(\ref{eqn:denom}) as a function of the total momentum $\vQ$ are plotted in Fig.~\ref{fig:mvals} for a range of parameters.
The diagonal wavevector $\vQ=Q_m(1,1)$ for $Q_m=1.19381$ point is very consistently the global minimum for a wide range of interaction parameters. Increasing either $J$ or $V_{pp}$ both have the effect of decreasing the minimum eigenvalue. However, larger $J$ tends to localize the minimum at $Q_m(1,1)$ whereas larger $V_{pp}$ has the opposite effect. A ridge of local minima extending down to the axial wavevector $\vQ=Q_m(1,0),\,Q_m(0,1)$ is also consistently present. Motivated by experiment, we discuss the order parameters associated with these wavevectors as well.

Some of the eigenvectors corresponding to both the diagonal and axial momenta are given in Table \ref{tab:evecs}. The $d$-wave character of the order parameter is somewhat harder to read off than in a one band model. As expected, for $\vQ=Q_m(1,1)$, both the on-site Copper amplitude ($l=1$) and the extended $s$-wave symmetry ($l=21$) vanish. For all three vectors presented, the weight is split primarily between the $l=2,3$ (on-site O $p_x$ and O $p_y$ amplitudes) and the $l=20$ (Cu-Cu $d$-form factor) basis vectors. Further, the $l=2$ and $l=3$ components are of the same order of magnitude and have opposite sign, indicating that these vectors are in fact primarily $d$-wave.

At $\vQ=Q_m(1,0)$, the order parameter is similarly primarily $d$-wave in character, though the $s$ and $s'$ components no longer vanish. As $V_{pp}$ is increased, the $d$-wave character also increases.

\begin{figure}[p!]
\includegraphics[scale=1.0]{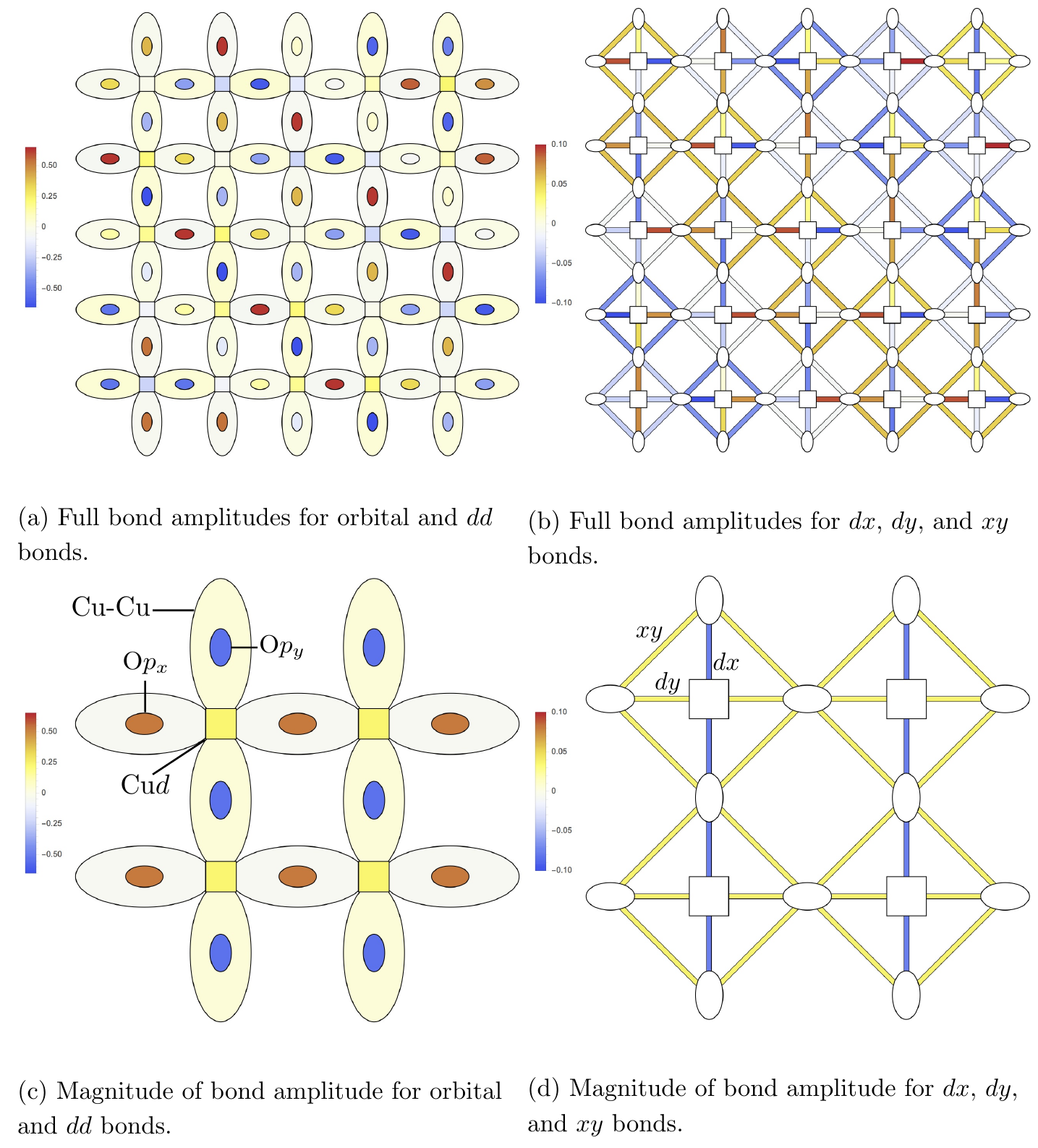}
\caption{(Color online) Real space representation of hopping amplitudes for diagonal $\vQ=Q_m(1,1)$ evaluated with $J=0.5$, $V_{pp}=1.25$, and $V_{pd}$ at $T=0.015$ and $p=0.1643$. For clarity the lattice has been divided into two separate pictures. (a) and (c) display the on-site Copper, the on-site oxygen and the copper-copper hopping amplitudes whereas (b) and (d) give the $dx$, $dy$ and $xy$ bond amplitudes. (a) and (b) plot the full functions given in Table \ref{tab:trsamplitudes} while (c) and (d) simply display the $\vr=0$ part. (c) and (d) also indicate which orbitals and bonds the symbols represent. The modulations of the order parameter in the diagonal direction can be seen in (a) and (b) by tracking the colors of the O $p_x$, O $p_y$ orbitals and the $xy$ bonds respectively over several unit cells. The (c) and (d) pictures make the $d$-form factor evident.}
\label{fig:LatPicDiag}
\end{figure}

\begin{figure}[p!]
\centering
\includegraphics[scale=1.0]{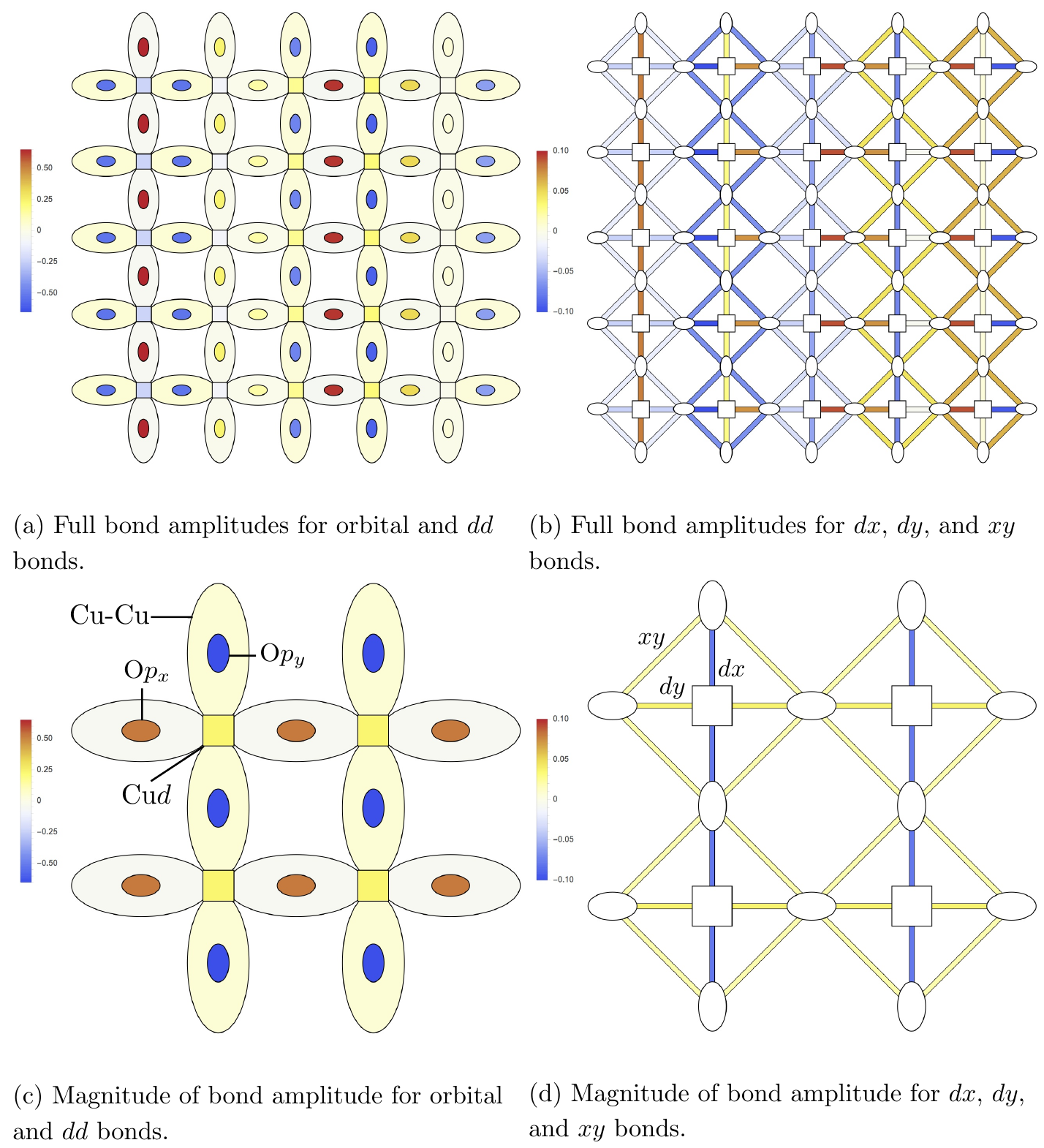}
\caption{(Color online) Just as in Fig.~\ref{fig:LatPicDiag}. 
Real space representation of hopping amplitudes for axial $\vQ=Q_m(1,0)$ evaluated with $J=0.5$, $V_{pp}=1.25$, and $V_{pd}$ at $T=0.015$ and $p=0.1643$. }
\label{fig:LatPicAx}
\end{figure}

Figs.~\ref{fig:LatPicDiag} and \ref{fig:LatPicAx} are visualizations of the amplitudes given by the order parameter $P^{\m\n}_\vQ(\vk)$ at $\vQ=Q_m(1,1)$ and $\vQ=Q_m(1,0)$ respectively. They are generated by taking the functions listed in Table \ref{tab:trsamplitudes} of Appendix~\ref{app:timerev} and plotting a corresponding color. Both with and without the spatial modulation $\sim\cos\vQ\cdot\vr$ envelope are shown for clarity. The primary difference between the two figures is the direction of the amplitude modulation shown in (a) and (b) of either figure. Further, technically, the difference in the colors representing the amplitudes of the O $p_x$ and O $p_y$ orbitals for the eigenvector at axial momentum in Fig.~\ref{fig:LatPicAx}(c) is not as strong as for the diagonal momentum in Fig.~\ref{fig:LatPicDiag}(c). This follows since, as mentioned, the $s$ and $s'$ components of the order parameter in the axial case are nonzero. However, since these components are small, there is little indication of any $s$ character.
\begin{table}
\centering
\begin{tabular}{r |c | l l l | l l l }
\multirow{3}{*}{$l$} & \multirow{3}{*}{$\phi^l_{\m\n}(\vk)$} & \multicolumn{3}{c|}{$Q_m(1,1)$} & \multicolumn{3}{c}{$Q_m(1,0)$} \\\cline{3-8}
& &\multicolumn{3}{c|}{} & \multicolumn{3}{c}{}  \\[-1.5ex]
& & \multicolumn{1}{c}{$V_{pp}=1.0$} & \multicolumn{1}{c}{$V_{pp}=1.5$} & \multicolumn{1}{c|}{$V_{pp}=2.0$}& \multicolumn{1}{c}{$V_{pp}=1.0$}& \multicolumn{1}{c}{$V_{pp}=1.5$} & \multicolumn{1}{c}{$V_{pp}=2.0$} \ttnl \hline\hline
& &\multicolumn{3}{c|}{} & \multicolumn{3}{c}{}  \\[-1.5ex]
$1$ &$\delta_{\mu d}\,\delta_{\nu d}$ & $\ph{-}0.0$ & $\ph{-}0.0$ & $\ph{-}0.0$ & $-0.3417$ & $-0.2348$ & $-0.1848$ \ttnl 
$2$ &$\delta_{\mu x}\,\delta_{\nu x}$ & $-0.4636$ & $-0.6185$ & $-0.6592$ & $-0.5252$ & $-0.6361$ & $-0.6659$\ttnl 
$3$ &$\delta_{\mu y}\,\delta_{\nu y}$  & $\ph{-}0.4636$ & $\ph{-}0.6185$ & $\ph{-}0.6592$ & $\ph{-}0.5426$ & $\ph{-}0.6500$ & $\ph{-}0.6756$ \ttnl  \hline
$4$ &$\delta_{\mu d}\,\delta_{\nu x}\sqrt{2}\cos\left({k_x\o2}\right)$  & $-0.2017$ & $-0.1184$ & $-0.0789$ & $-0.2065$ & $-0.1135$ & $-0.0751$ \ttnl  
$5$ &$\delta_{\mu x}\,\delta_{\nu d}\sqrt{2}\cos\left({k_x\o2}\right)$ & $\ph{-}0.2017$ & $\ph{-}0.1184$ & $\ph{-}0.0789$ & $\ph{-}0.2065$ & $\ph{-}0.1135$ & $\ph{-}0.0751$ \ttnl 
$6$ &$\delta_{\mu d}\,\delta_{\nu x}\sqrt{2}\sin\left({k_x\o2}\right)$ & $-0.2301$ & $-0.1374$ & $-0.0927$ & $-0.1618$ & $-0.0917$ & $-0.0612$ \ttnl  
$7$ &$\delta_{\mu x}\,\delta_{\nu d}\sqrt{2}\sin\left({k_x\o2}\right)$ & $-0.2301$ & $-0.1374$ & $-0.0927$ & $-0.1618$ & $-0.0917$ & $-0.0612$ \ttnl  
$8$ &$\delta_{\mu d}\,\delta_{\nu y}\sqrt{2}\cos\left({k_y\o2}\right)$ & $-0.2017$ & $-0.1184$ & $-0.0789$ & $\ph{-}0.0$ & $\ph{-}0.0$ & $\ph{-}0.0$ \ttnl 
$9$ &$\delta_{\mu y}\,\delta_{\nu d}\sqrt{2}\cos\left({k_y\o2}\right)$ & $\ph{-}0.2017$ & $\ph{-}0.1184$ & $\ph{-}0.0789$ & $\ph{-}0.0$ & $\ph{-}0.0$ & $\ph{-}0.0$ \ttnl  
$10$ &$\delta_{\mu d}\,\delta_{\nu y}\sqrt{2}\sin\left({k_y\o2}\right)$ & $-0.2301$ & $-0.1374$ & $-0.0927$ & $-0.1985$ & $-0.1154$ & $-0.0789$ \ttnl  
$11$ &$\delta_{\mu y}\,\delta_{\nu d}\sqrt{2}\sin\left({k_y\o2}\right)$ & $-0.2301$ & $-0.1374$ & $-0.0927$ & $-0.1985$ & $-0.1154$ & $-0.0789$\ttnl \hline 
$12$ &$\delta_{\mu x}\,\delta_{\nu y}\,2\cos\left({k_x\o2}\right)\cos\left({k_y\o2}\right)$ & $\ph{-}0.0$ & $\ph{-}0.0$ & $\ph{-}0.0$ & $\ph{-}0.0$ & $\ph{-}0.0$ & $\ph{-}0.0$ \ttnl  
$13$ &$\delta_{\mu y}\,\delta_{\nu x}\,2\cos\left({k_x\o2}\right)\cos\left({k_y\o2}\right)$ & $\ph{-}0.0$ & $\ph{-}0.0$ & $\ph{-}0.0$ & $\ph{-}0.0$ & $\ph{-}0.0$ & $\ph{-}0.0$ \ttnl  
$14$ &$\delta_{\mu x}\,\delta_{\nu y}\,2\cos\left({k_x\o2}\right)\sin\left({k_y\o2}\right)$ & $-0.1461$ & $-0.1334$ & $-0.1213$ & $-0.1595$ & $-0.1377$ & $-0.1242$ \ttnl  
$15$ &$\delta_{\mu y}\,\delta_{\nu x}\,2\cos\left({k_x\o2}\right)\sin\left({k_y\o2}\right)$ & $\ph{-}0.1461$ & $\ph{-}0.1334$ & $\ph{-}0.1213$ & $\ph{-}0.1595$ & $\ph{-}0.1377$ & $\ph{-}0.1242$ \ttnl  
$16$ &$\delta_{\mu x}\,\delta_{\nu y}\,2\sin\left({k_x\o2}\right)\cos\left({k_y\o2}\right)$ & $-0.1461$ & $-0.1334$ & $-0.1213$ & $\ph{-}0.0$ & $\ph{-}0.0$ & $\ph{-}0.0$ \ttnl 
$17$ &$\delta_{\mu y}\,\delta_{\nu x}\,2\sin\left({k_x\o2}\right)\cos\left({k_y\o2}\right)$ & $\ph{-}0.1461$ & $\ph{-}0.1334$ & $\ph{-}0.1213$ & $\ph{-}0.0$ & $\ph{-}0.0$ & $\ph{-}0.0$ \ttnl  
$18$ &$\delta_{\mu x}\,\delta_{\nu y}\,2\sin\left({k_x\o2}\right)\sin\left({k_y\o2}\right)$ & $\ph{-}0.0$ & $\ph{-}0.0$ & $\ph{-}0.0$ & $-0.0005$ & $-0.0004$ & $-0.0001$ \ttnl 
$19$ &$\delta_{\mu y}\,\delta_{\nu x}\,2\sin\left({k_x\o2}\right)\sin\left({k_y\o2}\right)$ & $\ph{-}0.0$ & $\ph{-}0.0$ & $\ph{-}0.0$ & $-0.0005$ & $-0.0004$ & $-0.0001$ \ttnl \hline
$20$ &$\delta_{\mu d}\,\delta_{\nu d}\left(\cos{k_x}-\cos k_y \right)$ & $\ph{-}0.3320$ & $\ph{-}0.1793$ & $\ph{-}0.1130$ & $\ph{-}0.2129$ & $\ph{-}0.1025$ & $\ph{-}0.0624$ \ttnl  
$21$ &$\delta_{\mu d}\,\delta_{\nu d}\left(\cos{k_x}+\cos k_y \right)$ & $\ph{-}0.0$ & $\ph{-}0.0$ & $\ph{-}0.0$ & $\ph{-}0.0209$ & $\ph{-}0.0112$ & $\ph{-}0.0072$ \ttnl 
$22$ &$\delta_{\mu d}\,\delta_{\nu d}\left(\sin{k_x}-\sin k_y \right)$ & $\ph{-}0.0$ & $\ph{-}0.0$ & $\ph{-}0.0$ & $\ph{-}0.0$ & $\ph{-}0.0$ & $\ph{-}0.0$ \ttnl 
$23$ &$\delta_{\mu d}\,\delta_{\nu d}\left(\sin{k_x}+\sin k_y \right)$ & $\ph{-}0.0$ & $\ph{-}0.0$ & $\ph{-}0.0$ & $\ph{-}0.0$ & $\ph{-}0.0$ & $\ph{-}0.0$ \ttnl 
\end{tabular}
\caption{Eigenvectors corresponding to $\vQ=Q_m(1,1)$  and $\vQ=Q_m(1,0)$ for $J=0.5$, $V_{pd}=1.0$ and $V_{pp}=1.0,\,1.5$ and $2.0$. The temperature is $T=0.015$ and the filling $p=0.1643$. All others parameters are as given in Table \ref{tab:param}.
}\label{tab:evecs}
\end{table}


The general structure of the pictures in Figs.~\ref{fig:LatPicDiag} and~\ref{fig:LatPicAx} are similar to those
obtained in Ref.~\onlinecite{SSRP13} for the one-band model. In the one-band case, the only quantities available
were the on-site densities on the Cu sites, and the bond order parameters involving nearest-neighbor pairs of Cu sites.
We find very similar modulations in the same quantities here. However, we now also have additional information using the O sites:
the on-site densities on the O sites, the bond orders between the nearest-neighbor Cu and O sites, and the diagonal bond orders between 
pairs of O sites. All of these quantities are also shown in the figures, and their spatial pattern mirrors those of Cu site and bond orders. 
In particular the modulation on an O site reflects that on the Cu-Cu bond it resides on.

\section{Small Fermi surfaces with antiferromagnetic order}
\label{sec:sfs}

We next consider the three band model in the presence of a staggered magnetic field pointing in the $\mathbf{\hat{x}}$-direction:
\eq{\label{eqn:hamHopAF}
\hat{{H}}_t'&=\ham_t + \ham_{AF} 
}
We perform a self-consistent Hartree-Fock analysis in order to determine the static antiferromagnetic (AF) order parameter $M_d$ on the copper atom sites. Further, the Coulomb interactions between the copper and oxygen orbitals may induce an antiferromagnetic bond order so that additional AF order parameters, $M_{pd}$ and $M_{pp}$, are required. It follows that the general extension to the hopping Hamiltonian in Eq.~(\ref{eqn:kernel}) is
\eq{\label{eqn:hamAF}
\ham_{AF}=-\sum_i e^{i\vK\cdot\vr_i}\s^x_{\a\b}\bigg[ &M_dc^\dag_{d\a}(\vr_i)c_{d\b}(\vr_i) \\
&+M_{pd}\big(-c_{d\a}^\dag(\vr)c_{x\b}(\vr)+c_{d\a}^\dag(\vr) c_{x\b}(\vr-\hat{\v{x}}) + 
c_{d\a}^\dag(\vr)c_{y\b}(\vr)-c_{d\a}^\dag (\vr)c_{y\b}(\vr-\hat{\v{y}})  + h.c. \big) \ph{\bigg]}\nt\\
&+M_{pp}\big(\, c_{x\a}^\dag(\vr) c_{y\b}(\vr)- c_{x\a}^\dag(\vr) c_{y\b}(\vr-\hat{\v{y}})\nt\\
&\ph{+M_{pp}\big( }-c_{x\a}^\dag(\vr-\hat{\v{x}}) c_{y\b}(\vr)+c_{x\a}^\dag(\vr-\hat{\v{x}})c_{y\b}(\vr-\hat{\v{y}}) +h.c.\big)\bigg] \nt
}
where $\vK=(\pi,\pi)$. The sign of the inter-orbital correlations is the same as in the original hopping Hamiltonian (see Figs.~\ref{fig:hoppingintparams} and \ref{fig:amplitudes}).
We transform this Hamiltonian to momentum space basis functions in Appendix~\ref{app:basisAF}, and describe how the magnetic order parameters
$M_d$, $M_{pd}$, and $M_{pp}$ are computed in the Hartree-Fock theory. We find that $M_{pd}$ and $M_{pp}$ have near-vanishing magnitude and they will not be discussed further.

The particle-hole $T$-matrix calculation in the presence of AF order is similar to the one presented in Section \ref{sec:Tmatrix}, though considerably more complicated 
due to spin-flip processes. These calculations are presented in Appendix \ref{app:AFTmatrix}. 
While we are still primarily interested in the particle-hole spin singlet channel, the presence of AF order breaks the SU(2) symmetry of the original Hamiltonian causing the charge channel at wavevector $\vQ$ to mix with the spin channel at wavevector $\vQ+(\pi, \pi)$. However, while the total spin is no longer conserved, the $S_x$ component still is and mixing only occurs between the particle-hole pair with total spin $S=0$ and the particle-hole pair with total spin $S=1$ and spin component $S_x = 0$. 
Having to track the total spin doubles the number of required basis functions so that the order parameter is a 46-component vector $\{\P_l(\vQ)\}$.  
This analysis, as well as the basis functions used for the actual calculations, are given in Appendix \ref{app:symmetries}. An additional inversion symmetry is present, but 
instead of being used to decrease the number of basis functions, it was used to verify our results.
\subsection{Results}
\begin{figure}[p!]
\centering
\includegraphics[scale=1.0]{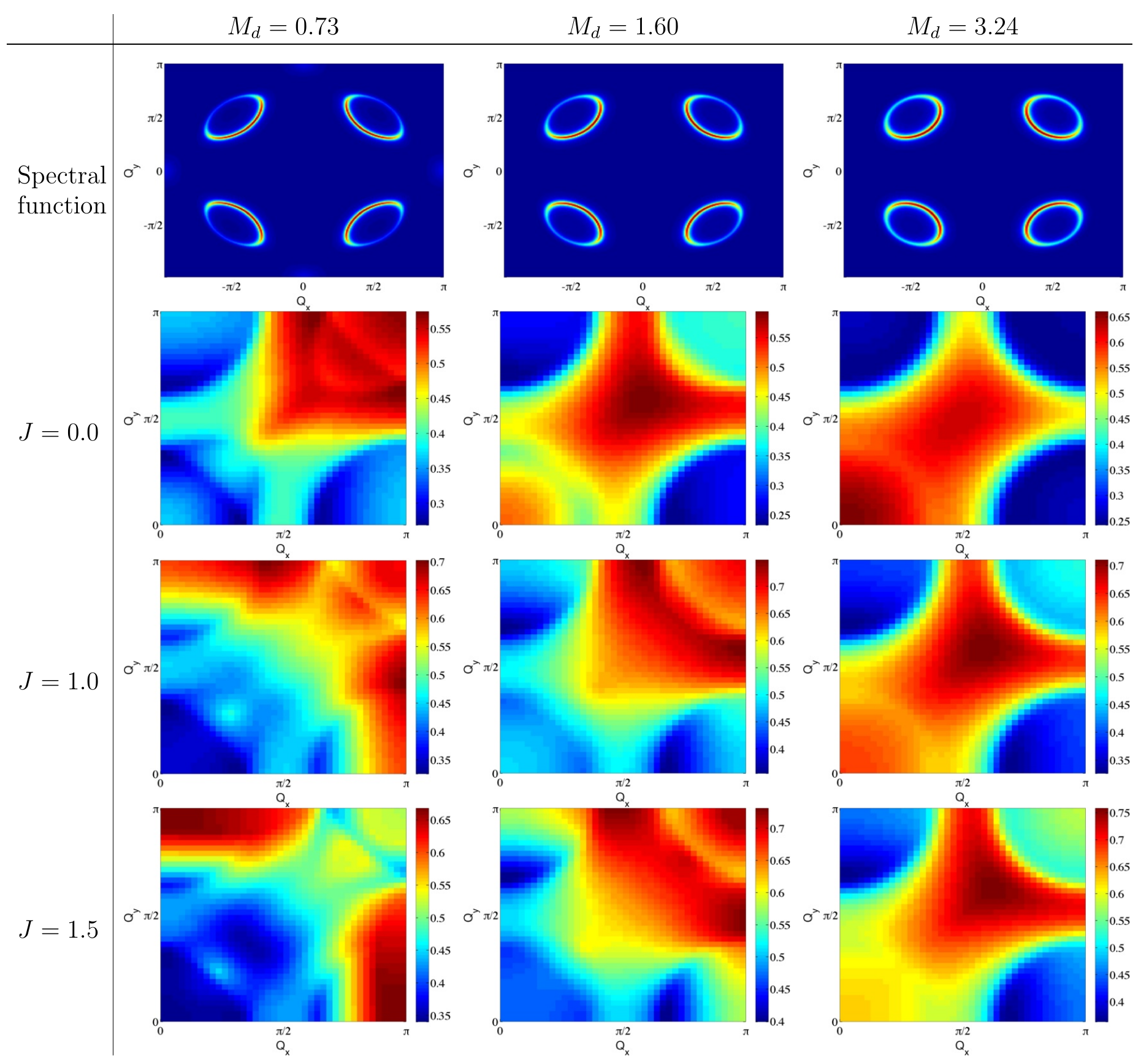}
\caption{(Color online) Spectral functions and minimum eigenvalues for $M_d=$0.73, 1.60 and 3.24 (which corresponds to $U_d+2J=$ 3.25, 5.0 and 8.0 respectively). The chemical potential is adjusted so that $p\sim0.10$, while $V_{pp}=1.5$ and $V_{pd}=1.0$. The second through fourth columns are for $J=0.0,\,1.0,$ and $1.5$ respectively. 
Note that the minimum eigenvalues are mostly at $(Q_2, 0)$ with $Q_2 \approx 2 \pi/3$; in some cases there are also well-formed minima 
at $(Q_1, 0)$ with $Q_1 \approx \pi/3$.
}\label{HFFull2}
\end{figure}

Fig.~\ref{HFFull2} shows the spectral functions and minimum eigenvalues for $U_d+2J$ ranging from 3.25 to 8.0 with the chemical potential chosen so that the hole density $p\sim0.10$. The other parameters are 
given in the bottom row of Table \ref{tab:param}.

As is apparent from Fig.~\ref{HFFull2}, the minimal eigenvalues are consistently along the axes either at $(\pm Q_{1,2}, 0)$ and $(0, \pm Q_{1,2})$, with 
$Q_1 \approx \pi/3$ and $Q_2 \approx 2 \pi/3$. The orientation of the eigenvalue is therefore in accord with experiments. The global minimum is mostly at the wavevector $Q_2$, 
which corresponds 
approximately to the distance between the tips of the hole pockets shown in the top row of Fig.~\ref{HFFull2}. 
In a few cases, there is also a well-formed minimum
at $Q_1$; we do not have a correspondingly simple interpretation
of $Q_1$, but suspect that it is related to the incipient electron pocket near the antinodes which is present at smaller magnetic order. 

\begin{figure}
\includegraphics[scale=1.0]{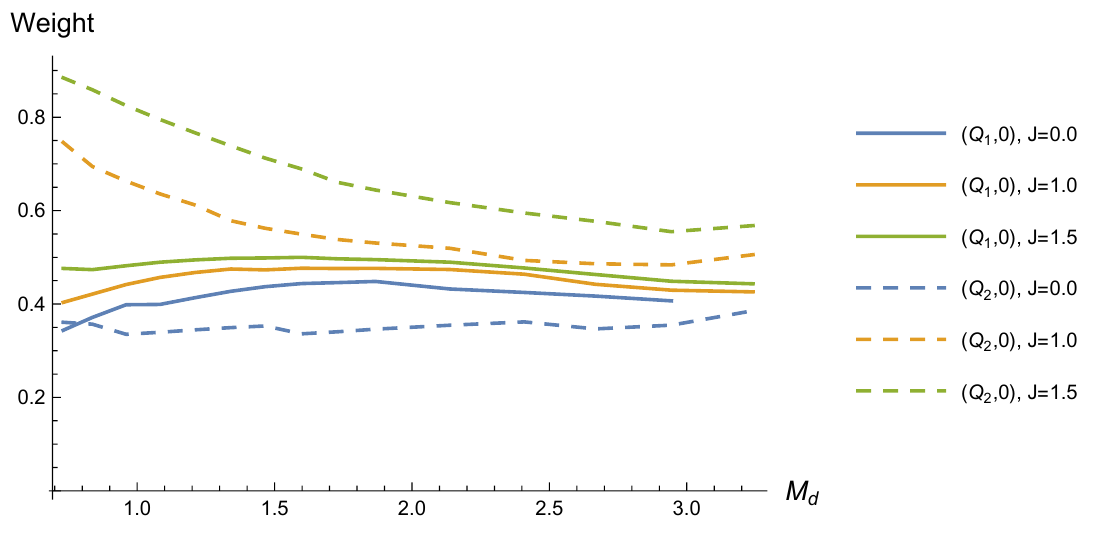}
\caption{(Color online) The sum of the squares of the $S=0$ components of the original eigenvectors corresponding to the minimum eigenvalues at $(Q_1,0)$ and $(Q_2,0)$. That is, we plot $\sum_{l=1}^{23}\left|\P_l(\vQ)\right|^2$ at $\vQ=(Q_{1,2},0)$ where $\{\P_l(\vQ)\}$ are the components corresponding to the 46 basis functions given in Tables~\ref{tab:basisAFX2p1} and \ref{tab:basisAFX2p2}. We subsequently project out the $S=1$ components ($l=24-46$) and normalize.}\label{fig:singweight}
\end{figure}

Turning to the form factors, recall our observation above that the eigenmodes have components both in the $S=0$ charge density wave at $\vQ$ and in the
$S=1$ spin density wave at $\vQ + (\pi, \pi)$. We show in Fig.~\ref{fig:singweight} the relative weights of the $S=0$ and $S=1$ components 
at the wavevectors $(Q_1, 0)$ and $(Q_2, 0)$. For most of the cases, the weight in the spin density wave component is actually dominant. 
This appears to be due to the proximity of the critical point where the antiferromagnetic order at $(\pi, \pi)$ vanishes, and amplitude fluctuations
in the N\'eel order are enhanced. 

\begin{figure}[p!]
\includegraphics[scale=1.0]{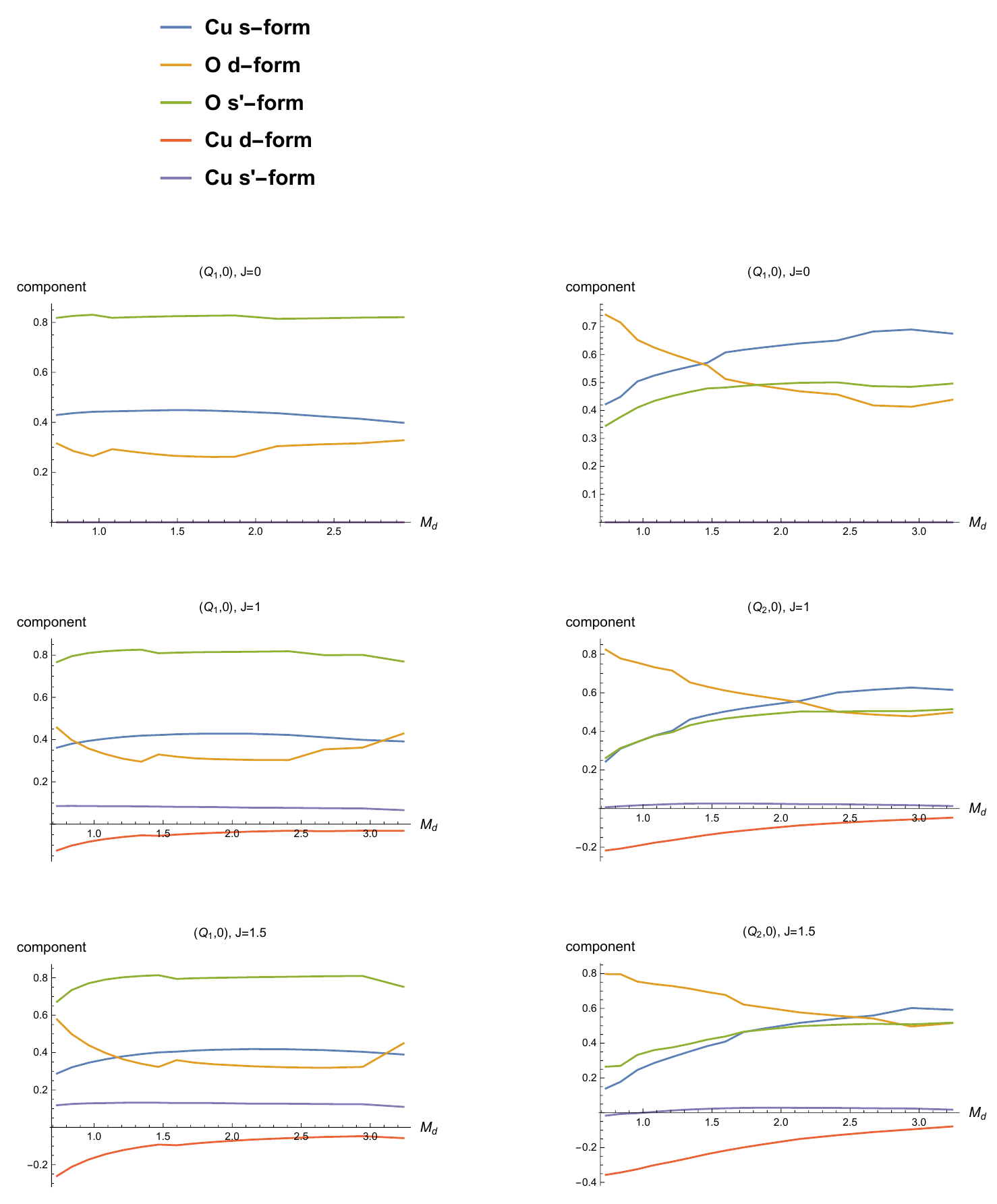}
\caption{(Color online) $S=0$ form factor dependence at $(Q_1 ,0)$ and $(Q_2 ,0)$, as a function of magnetization $M_d$ (from which $U_d+2J$ is determined). These values are determined by projecting out the $S=1$ components of the eigenvectors and subsequently normalizing. Only the most important contributions are displayed. On the first row, the legend is shown. In descending order the remaining rows show the results for $J=0.0,\,1.0$ and $1.5$. The other interaction parameters are given in Table~\ref{tab:param} and the chemical potential is chosen so that $p\sim0.10$.\label{fig:formMd}}
\end{figure}

Nonetheless, when we take into account the orientational fluctuations of the N\'eel order induced by a non-zero temperature, we expect that the $S=1$ components
will average to zero . 
For this reason, we focus on the spatial structure of the $S=0$ component of the order parameter alone. 
The normalized components of the eigenvector projected into the $S=0$ components are shown in Fig.~\ref{fig:formMd}.
The consistent trend in these plots, and one of our key results, is that increasing the magnetic order, $M_d$, leads to a decrease in the $d$ components
and a corresponding increase in the $s'$ components. 


\section{Conclusions}

This paper has analyzed charge ordering instabilities of 3-band models of the cuprates. Consistent with earlier results on related models,
we find that starting from a metal with a large Fermi surface invariably leads, in the simplest RPA approximation, to charge-ordering
along a `diagonal' wavevector, which disagrees with experimental observations. However, as suggested in Ref.~\onlinecite{3band}, 
starting from a Fermi surface reconstructed by antiferromagnetic order leads to the observed charge ordering along the principal axes.
We examined the form-factors of this ordering, and found that its $d$-wave character was suppressed as the strength of the magnetic order
was increased. This trend is consistent with recent X-ray experimental observations of charge order in LBCO in Ref.~\onlinecite{Achkar14},
which measured the ratio of $s'$ to $d$ components on the O sites.
Our results for these parameters are in Fig.~\ref{fig:formMd}. The magnetically ordered LBCO compound has a much larger $s'/d$ ratio
than that observed by STM in the non-magnetic compounds \cite{Fujita2014}.

The model of magnetic order used in the present paper is rather crude, and it would be interesting to extend the computations to more 
realistic models. We have assumed magnetic order at $(\pi, \pi)$, whereas the magnetic order in LBCO is incommensurate. 
The magnetic order has been assumed to be static, but it would be interesting to examine the influence of a frequency-dependent 
electronic self energy in a Eliashberg framework. This would then complement the spin-liquid perspective taken recently in Ref.~\onlinecite{DCSS14}.

Finally, we note that our renormalized classical treatment of magnetic order here is more appropriate 
for the electron-doped cuprates \cite{Tremblay04}. Interestingly, charge order has recently been observed
in an electron-doped compound \cite{edoped} with a wavevector which is close to the $(\pm Q_1, 0)$ wavevector
found in our computations above. It would be interesting to measure the form factor of this ordering: the implication of our results
here is that the $s$-form factor will be larger than in the hole-doped cuprates.

\section*{Acknowledgments} 

We thank B.~Atkinson, D.~Chowdhury, D.~Hawthorn, and A.~Kampf for useful discussions.  
This research was 
supported by the NSF under Grant DMR-1360789, the Templeton foundation, and MURI grant W911NF-14-1-0003 from ARO.
Research at Perimeter Institute is supported by the Government of Canada through Industry Canada 
and by the Province of Ontario through the Ministry of Research and Innovation.   \\

\appendix

\section{Basis functions} \label{app:timerev}

This appendix expresses the Hamiltonian in Eq.~(\ref{eqn:ham}) in Fourier space, and then writes it in terms of basis
functions which aid in the determination of the eigenmodes in the particle-hole sector.
We begin by introducing the Fourier transforms
\eq{\label{eqn:fourdef}
c_{d\a}(\vr_i)&=\sum_\vk e^{-i\vk\cdot \vr_i}c_{d\a}(\vk),
&
c^\dag_{d\a}(\vr_i)&=\sum_\vk e^{i\vk\cdot \vr_i}c^\dag_{d\a}(\vk) & &\\
c_{\mu\a}(\vr_i)&=-i\sum_\vk e^{-i\vk\cdot (\vr_i+\v{R}_\mu)}c_{\mu\a}(\vk),
&
c^\dag_{\mu\a}(\vr_i)&=i\sum_\vk e^{i\vk\cdot (\vr_i+\v{R}_\mu)}c^\dag_{\mu\a}(\vk)
&
\mu=x,y\nt }
where $\v{R}_\mu$ is the position within the unit cell of the $\mu$th orbital: $\v{R}_d=0$, $\v{R}_x=+ \hat{\v{x}}/2$, and $\v{R}_y=+ \hat{\v{y}}/2$. The Coulomb terms become
\eq{
\ham_C^h=\sum_{\vk,\vk',\vq}& \bigg[
	U_d\, c_{d\ua}^\dag(\vk'-\vq/2)c_{d\ua}(\vk-\vq/2)c_{d\da}^\dag(\vk+\vq/2)c_{d\da}(\vk'+\vq/2) \\
	&+ U_p\bigg( c_{x\ua}^\dag(\vk'-\vq/2)c_{x\ua}(\vk-\vq/2)c_{x\da}^\dag(\vk+\vq/2)c_{x\da}(\vk'+\vq/2) \nt\\
	&\ph{+ U_p\bigg( }+ c_{y\ua}^\dag(\vk'-\vq/2)c_{y\ua}(\vk-\vq/2)c_{y\da}^\dag(\vk+\vq/2)c_{y\da}(\vk'+\vq/2) \bigg)\bigg] \nt \\
\ham_C^v=\sum_{\vk,\vk',\vq}& \bigg[
	2V_{pd}\bigg(
	\cos\left({k_x-k_x'\o2}\right)c_{d\a}^\dag(\vk'-\vq/2)c_{d\a}(\vk-\vq/2)c_{x\b}^\dag(\vk+\vq/2)c_{x\b}(\vk'+\vq/2) \\
	&+ \cos\left({k_y-k_y'\o2}\right)c_{d\a}^\dag(\vk'-\vq/2)c_{d\a}(\vk-\vq/2)c_{y\b}^\dag(\vk+\vq/2)c_{y\b}(\vk'+\vq/2) 	\bigg)\nt \\
	&+ 4V_{pp}\cos\left({k_x-k_x'\o2}\right)\cos\left({k_y-k_y'\o2}\right) 
	c_{x\a}^\dag(\vk'-\vq/2)c_{x\a}(\vk-\vq/2)c_{y\b}^\dag(\vk+\vq/2)c_{y\b}(\vk'+\vq/2)
	\bigg]\nt}
and the copper-copper exchange interaction is given by
\eq{
\ham_J=\sum_{\vk,\vk',\vq}\sum_a{J\o 4}\left(\cos(k_x-k_x')+\cos(k_y-k_y')\right) c^\dag_{d\a}(\vk'-\vq/2)\s^a_{\a\b}c_{d\b}(\vk-\vq/2)c^\dag_{d\gamma}(\vk+\vq/2)\s^a_{\gamma\delta}c_{d\delta}(\vk'+\vq/2)\,.}
These expressions may be simplified by writing them as a sum over the basis functions $\phi^{\,l}_{\m\n}(\vk)$ given in Table \ref{tab:basis}. 
\begin{table}
\center
\begin{tabular}{c c}
\begin{minipage}{.5\linewidth}
\center
\begin{tabular}{r | c c | c }
$l$ & $\mu$ & $\nu$ &$\phi^{\,l}_{\mu\nu}(\vk)$ \tnl\hline\hline
1 & $d$ & $d$ &	$\delta_{\mu d}\,\delta_{\nu d}$   \tnl 
2 & $x$ & $x$ &	$\delta_{\mu x}\,\delta_{\nu x}$ \tnl 
3 & $y$ & $y$ &	$\delta_{\mu y}\,\delta_{\nu y}$ \tnl \hline
4 & $d$ & $x$ &	$\delta_{\mu d}\,\delta_{\nu x}\sqrt{2}\cos\left({k_x\o2}\right)$ \tnl
5 & $x$ & $d$ & 	$\delta_{\mu x}\,\delta_{\nu d}\sqrt{2}\cos\left({k_x\o2}\right)$ \tnl 
6 & $d$ & $x$ &	$\delta_{\mu d}\,\delta_{\nu x}\sqrt{2}\sin\left({k_x\o2}\right)$ \tnl
7 & $x$ & $d$ & 	$\delta_{\mu x}\,\delta_{\nu d}\sqrt{2}\sin\left({k_x\o2}\right)$ \tnl\hline
8 & $d$ & $y$ &	$\delta_{\mu d}\,\delta_{\nu y}\sqrt{2}\cos\left({k_y\o2}\right)$ \tnl
9 & $y$ & $d$ &	$\delta_{\mu y}\,\delta_{\nu d}\sqrt{2}\cos\left({k_y\o2}\right)$ \tnl 
10 & $d$ & $y$ &	$\delta_{\mu d}\,\delta_{\nu y}\sqrt{2}\sin\left({k_y\o2}\right)$ \tnl
11 & $y$ & $d$ &	$\delta_{\mu y}\,\delta_{\nu d}\sqrt{2}\sin\left({k_y\o2}\right)$\tnl
\multicolumn{2}{c}{ } & \multicolumn{2}{c}{ $\phantom{\delta_{\mu x}\,\delta_{\nu y}\,2\cos\left({k_x\o2}\right)\cos\left({k_y\o2}\right)}$ }
\end{tabular}
\end{minipage} 
&
\begin{minipage}{.5\linewidth}
\center
\begin{tabular}{r | c c | c }
$l$ & $\mu$ & $\nu$ &$\phi^{\,l}_{\mu\nu}(\vk)$ \tnl \hline\hline
12 & $x$ & $y$ &  	$\delta_{\mu x}\,\delta_{\nu y}\,2\cos\left({k_x\o2}\right)\cos\left({k_y\o2}\right)$   \tnl
13 & $y$ & $x$ &  	$\delta_{\mu y}\,\delta_{\nu x}\,2\cos\left({k_x\o2}\right)\cos\left({k_y\o2}\right)$  \tnl 
14 &  $x$ & $y$ & 	$\delta_{\mu x}\,\delta_{\nu y}\,2\cos\left({k_x\o2}\right)\sin\left({k_y\o2}\right)$  \tnl
15 & $y$ & $x$ & 	$\delta_{\mu y}\,\delta_{\nu x}\,2\cos\left({k_x\o2}\right)\sin\left({k_y\o2}\right)$ \tnl
16 &  $x$ & $y$ & 	$\delta_{\mu x}\,\delta_{\nu y}\,2\sin\left({k_x\o2}\right)\cos\left({k_y\o2}\right)$   \tnl
17 & $y$ & $x$ & 	$\delta_{\mu y}\,\delta_{\nu x}\,2\sin\left({k_x\o2}\right)\cos\left({k_y\o2}\right)$  \tnl
18 &  $x$ & $y$ & 	$\delta_{\mu x}\,\delta_{\nu y}\,2\sin\left({k_x\o2}\right)\sin\left({k_y\o2}\right)$\tnl
19 & $y$ & $x$ & 	$\delta_{\mu y}\,\delta_{\nu x}\,2\sin\left({k_x\o2}\right)\sin\left({k_y\o2}\right)$\tnl \hline
20 & $d$ & $d$ & 	$\delta_{\mu d}\,\delta_{\nu d}\left(\cos{k_x}-\cos k_y \right)$ \tnl
21 & $d$ & $d$ & 	$\delta_{\mu d}\,\delta_{\nu d}\left(\cos{k_x}+\cos k_y \right)$  \tnl
22 & $d$ & $d$ & 	$\delta_{\mu d}\,\delta_{\nu d}\left(\sin{k_x}-\sin k_y \right)$  \tnl
23 & $d$ & $d$ & 	$\delta_{\mu d}\,\delta_{\nu d}\left(\sin{k_x}+\sin k_y \right)$ 
\end{tabular}
\end{minipage}
\end{tabular}
\caption{For each $l$-index,  $\phi^l_{\mu\nu}(\vk)$ is nonzero only for the $\mu\nu-$pair given in the second and third columns of each table. The full function is shown in the fourth column.}\label{tab:basis}
\end{table}
In this basis, the interaction Hamiltonian becomes
\eq{
\ham_C+\ham_J=
\sum_{\vk,\vk',\vq}\bigg[ & \sum_{l=1}^{19}\sum_{\mu\nu} {\V_l\o2} \phi^{\,l}_{\mu\n}(\vk){\phi^{\,l}_{\m\n}(\vk')}  c^\dag_{\mu\a}(\vk'-\vq/2)c_{\mu\a}(\vk-\vq/2)c^\dag_{\nu\b}(\vk+\vq/2)c_{\nu\b}(\vk'+\vq/2)\\
	&+\sum_{l=20}^{23} \sum_{\mu\nu}{\V_l\o6} \phi^{\,l}_{\mu\nu}(\vk){\phi^{\,l}_{\m\n}(\vk')}  c^\dag_{d\a}(\vk'-\vq/2)\s^a_{\a\b}c_{d\b}(\vk-\vq/2)c^\dag_{d\gamma}(\vk+\vq/2)\s^a_{\gamma\delta}c_{d\delta}(\vk'+\vq/2) \bigg]\,.\nt
}
where the interaction parameters $\V_l$ are given by
\eq{
\V_l =
\left\{
\begin{matrix*}[l]
U_d, & l=1 \\
U_p, & l=2,\,3 \\
V_{pd}, & l=4-11 \\
V_{pp}, & l=12-19 \\
3J/4, & l=20-23
\end{matrix*}\right. \,.\label{eqn:intparam}}

The action of time-reversal on the basis functions is summarized in Table~\ref{tab:timerev}.
\begin{table}[h]
\center
\begin{tabular}{c c}
\begin{minipage}{.5\linewidth}
\center
\begin{tabular}{r | c c | c c c  }
$l$ & $\mu$ & $\nu$ &\multicolumn{3}{c}{$\T\P_l(\vQ)$} \tnl \hline\hline
1 & $d$ & $d$ &	$\T: \P_1(\vQ)$&$\mapsto$&	$\ph{-}\P_1(\vQ)$   \tnl
2 & $x$ & $x$ &	$\T: \P_2(\vQ)$&$\mapsto$&$\ph{-}\P_2(\vQ)$ \tnl
3 & $y$ & $y$ &	$\T: \P_3(\vQ)$&$\mapsto$&$\ph{-}\P_3(\vQ)$ \tnl \hline
4 & $d$ & $x$ &	$\T: \P_4(\vQ)$&$\mapsto$&$-\P_5(\vQ)$ \tnl
5 & $x$ & $d$ & 	$\T: \P_5(\vQ)$&$\mapsto$&$-\P_4(\vQ)$ \tnl 
6 & $d$ & $x$ &	$\T: \P_6(\vQ)$&$\mapsto$&$\ph{-}\P_7(\vQ)$ \tnl
7 & $x$ & $d$ & 	$\T: \P_7(\vQ)$&$\mapsto$&$\ph{-}\P_6(\vQ)$ \tnl\hline
8 & $d$ & $y$ &	$\T: \P_8(\vQ)$&$\mapsto$&$-\P_9(\vQ)$ \tnl
9 & $y$ & $d$ &	$\T: \P_9(\vQ)$&$\mapsto$&$-\P_8(\vQ)$ \tnl10 & $d$ & $y$ &	$\T: \P_{10}(\vQ)$&$\mapsto$&$\ph{-}\P_{11}(\vQ)$ \tnl
11 & $y$ & $d$ &	$\T: \P_{11}(\vQ)$&$\mapsto$&$\ph{-}\P_{10}(\vQ)$\tnl
\multicolumn{2}{c}{ } & \multicolumn{4}{c}{\phantom{ $\ph{-}\P_{10}(\vQ)$} }
\end{tabular}
\end{minipage} 
&
\begin{minipage}{.5\linewidth}
\center
\begin{tabular}{r | c c | c c c  }
$l$ & $\mu$ & $\nu$ &\multicolumn{3}{c}{$\T\P_l(\vQ)$} \tnl \hline\hline
12 & $x$ & $y$ &  	$\T: \P_{12}(\vQ)$&$\mapsto$&$\ph{-}\P_{13}(\vQ)$   \tnl
13 & $y$ & $x$ &  	$\T: \P_{13}(\vQ)$&$\mapsto$&$\ph{-}\P_{12}(\vQ)$  \tnl 
14 &  $x$ & $y$ & 	$\T: \P_{14}(\vQ)$&$\mapsto$&$-\P_{15}(\vQ)$  \tnl
15 & $y$ & $x$ & 	$\T: \P_{15}(\vQ)$&$\mapsto$&$-\P_{14}(\vQ)$ \tnl
16 &  $x$ & $y$ & 	$\T: \P_{16}(\vQ)$&$\mapsto$&$-\P_{17}(\vQ)$   \tnl
17 & $y$ & $x$ & 	$\T: \P_{17}(\vQ)$&$\mapsto$&$-\P_{16}(\vQ)$  \tnl
18 &  $x$ & $y$ & 	$\T: \P_{18}(\vQ)$&$\mapsto$&$\ph{-}\P_{19}(\vQ)$\tnl
19 & $y$ & $x$ & 	$\T: \P_{19}(\vQ)$&$\mapsto$&$\ph{-}\P_{18}(\vQ)$\tnl \hline
20 & $d$ & $d$ & 	$\T: \P_{20}(\vQ)$&$\mapsto$&$\ph{-}\P_{20}(\vQ)$ \tnl
21 & $d$ & $d$ & 	$\T: \P_{21}(\vQ)$&$\mapsto$&$\ph{-}\P_{21}(\vQ)$  \tnl
22 & $d$ & $d$ & 	$\T: \P_{22}(\vQ)$&$\mapsto$&$-\P_{22}(\vQ)$  \tnl
23 & $d$ & $d$ & 	$\T: \P_{23}(\vQ)$&$\mapsto$&$-\P_{23}(\vQ)$ 
\end{tabular}
\end{minipage}
\end{tabular}
\caption{The actions of time-reversal on the basis function coefficients $\{\P_l(\vk)\}$.}
\label{tab:timerev}
\end{table}

\begin{figure}\centering
\includegraphics[scale=1.0]{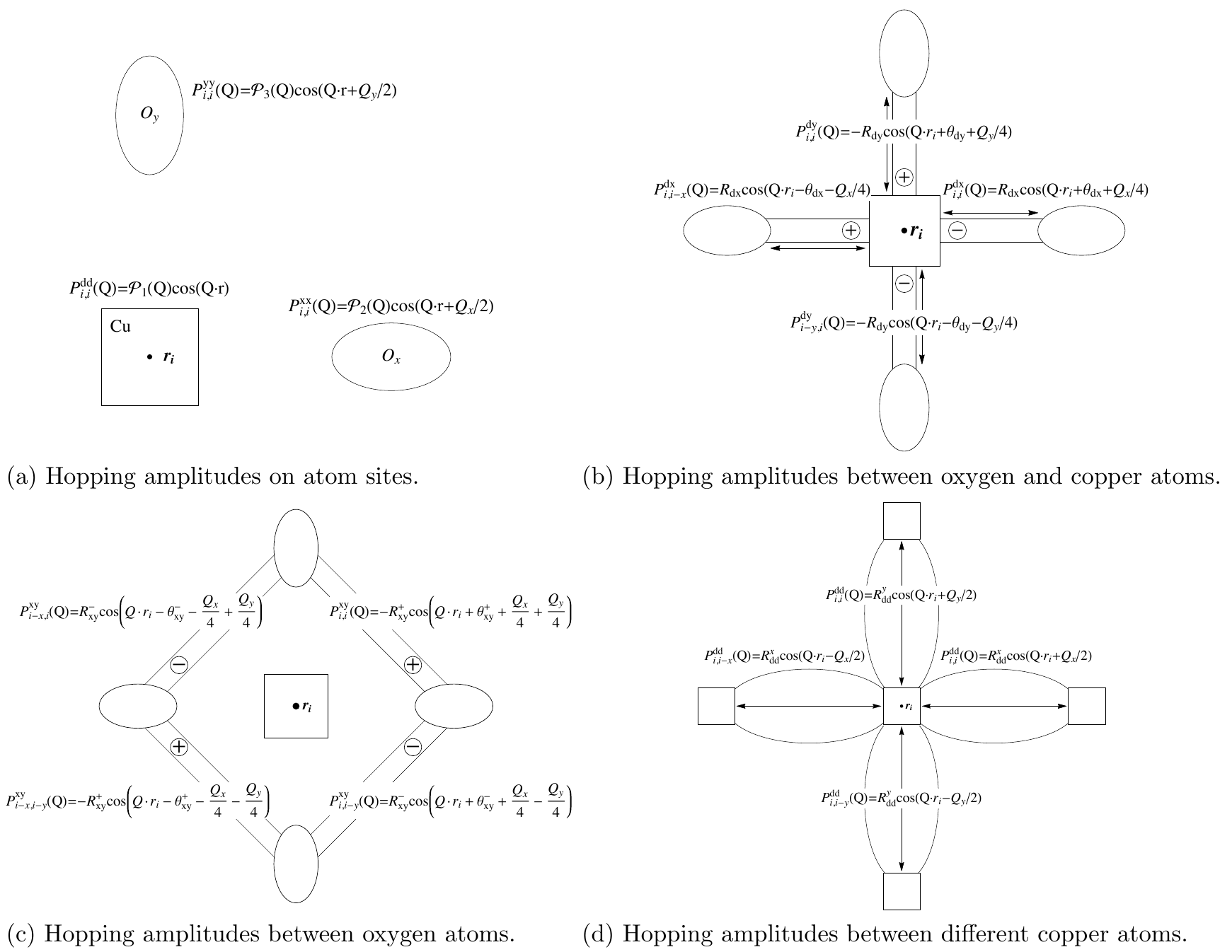}
\caption{Real space representation of hopping amplitudes. The signs written in the $dx$, $dy$, and $xy$ bonds in (b) and (c) indicate which sign the hopping term in the Hamiltonian corresponding to that bond has. The amplitudes given in Table \ref{tab:trsamplitudes} have been multiplied by this phase.}
\label{fig:amplitudes}
\end{figure}

Since the eigenvectors corresponding to the lowest eigenvalues are in general time-reversal preserving, we focus on this case. Table \ref{tab:trsamplitudes} summarizes the relationship between the real-space order parameter $P_{ij}^{\m\n}$ and an eigenvector $\{\P_l(\vQ)\}$. Note that the amplitude is multiplied by the sign of the hopping term in the Hamiltonian corresponding to that bond has. Fig.~\ref{fig:amplitudes} gives these signs and shows how these look on the lattice.
\begin{table}
\centering
\begin{tabular}{c | r c l | l  }
Bond &\multicolumn{3}{c|}{Definitions} & \multicolumn{1}{c}{$P_{ij}^{\m\n}$}  \\ \hline\hline
 & & & & $\P_1\cos\(\vQ\cdot\vr_i\)\d_{i,j}$  \\
\multirow{2}{*}{$dd$}& $R_{dd}^x$&= &$\(-\P_{20}+\P_{21}\)/2$ & $+R^x_{dd}\,\big[\cos\(\vQ\cdot\vr_i+{Q_x\o2}\)\d_{i,j-\hat{\v{x}}}$\\ 
& $R_{dd}^y$&= &$\(\P_{20}+\P_{21}\)/2$ & $\ph{R_{dd}^x\big[}+\cos\(\vQ\cdot\vr_i-{Q_x\o2}\)\d_{i,j+\hat{\v{x}}}\big]$\\ 
& & & & $+R_{dd}^y\big[\cos\(\vQ\cdot\vr_i+{Q_y\o2}\)\d_{i,j-\hat{\v{y}}}$\\ 
& & & & $\ph{R^y_{dd}\big[}+\cos\(\vQ\cdot\vr_i-{Q_y\o2}\)\d_{i,j+\hat{\v{y}}}\big]$\\ \hline
$xx$ & & & & $\P_2\cos\(\vQ\cdot\vr_i+{Q_x\o2}\)\d_{i,j}$ \\\hline
$yy$ & & &  &$\P_3\cos\(\vQ\cdot\vr_i+{Q_y\o2}\)\d_{i,j}$\\\hline
\multirow{2}{*}{$dx$} & $R_{dx}$&$=$&$\sqrt{\(\P_4^2+\P_6^2\)/2}$ &${R_{dx}}\Big[ \cos\(\vQ\cdot\vr_i+{Q_x\o4}+\theta_{dx}\)\d_{i,j}$\\
& $\theta_{dx}$&$=$& $\arctan\({\P_4/\P_6}\)$ & $\ph{R2}+\cos\(\vQ\cdot\vr_i-{Q_x\o4}-\theta_{dx}\)\delta_{i,j+\hat{\v{x}}}\Big]$ \\ \hline
\multirow{2}{*}{$dy$} & $R_{dy}$&$=$& $\sqrt{\(\P_8^2+\P_{10}^2\)/2}$ &${R_{dy}}\Big[ -\cos\(\vQ\cdot\vr_i+{Q_y\o4}+\theta_{dy}\)\d_{i,j}$\\
& $\theta_{dy}$&$=$& $\arctan\({\P_8/\P_{10}}\)$ & $\ph{R_{dy}\big[}-\cos\(\vQ\cdot\vr_i-{Q_y\o4}-\theta_{dy}\)\d_{i,j+\hat{\v{y}}}\Big]$\\ \hline
\multirow{4}{*}{$xy$} & $R_{xy}^-$&$=$&${1\o2}\sqrt{\(\P_{12}-\P_{18}\)^2+\(\P_{14}+\P_{16}\)^2}$ & 
$+{R_{xy}^-}\Big[\cos\(\vQ\cdot\vr_i +{Q_x\o4}-{Q_y\o4}+\theta_{xy}^-\)\d_{i,j+\hat{\v{y}}}$ \\
& $\theta_{xy}^-$&$=$& $\arctan\big[\(\P_{14}+\P_{16}\)/\(\P_{12}-\P_{18}\)\big]$&
$\ph{{R_{xy}^-\o2}}\,+\cos\(\vQ\cdot\vr_i +{3Q_x\o4}+{Q_y\o4}-\theta_{xy}^-\)\d_{i,j-\hat{\v{x}}}\Big]$\\
& $R_{xy}^+$&$=$&${1\o2}\sqrt{\(\P_{12}+\P_{18}\)^2+\(-\P_{14}+\P_{16}\)^2}$ &
$-{R_{xy}^+}\Big[\cos\(\vQ\cdot\vr_i +{Q_x\o4}+{Q_y\o4}+\theta_{xy}^+\)\d_{i,j}$ \\
& $\theta_{xy}^+$&$=$&$\arctan\big[\(-\P_{14}+\P_{16}\)/\(\P_{12}+\P_{18}\)\big]$ &
$\ph{{R_{xy}^-\o2}}\,+\cos\(\vQ\cdot\vr_i+{3Q_x\o4}-{Q_y\o4}-\theta_{xy}^+\)\d_{i,j-\hat{\v{x}}+\hat{\v{y}}}\Big]$ 
\end{tabular}
\caption{Transition amplitudes in the time-reversal invariant sector at wave vector $\vQ$.}
\label{tab:trsamplitudes}
\end{table}

\section{Basis functions with antiferromagnetic order}
\label{app:basisAF}

To determine the momentum space representation of the Hamiltonian in the presence of antiferromagnetic order in 
Eq.~(\ref{eqn:hamAF}), we begin by introducing a new electron operator $\psi_{\m\s}$ :
\eq{
\psi_{\m\ua}(\vk)&=c_{\m\ua}(\vk) & \psi_{\m\da}(\vk)&=c_{\m\da}(\vk+\vK)\,.
}
With these operators we re-express Eq.~\ref{eqn:hamAF} as
\eq{
\ham_{AF}=-\sum_\vk \Bigg[ &M_d\( \psi_{d\ua}^\dag(\vk)\psi_{d\da}(\vk) + h.c. \) \\
&+2M_{pd} \Big( -\cos\({k_x\o2}\) \psi_{d\ua}^\dag(\vk)\psi_{x\da}(\vk+\vK)+\sin\({k_x\o2}\)\psi_{x\ua}^\dag(\vk)\psi_{d\da}(\vk) +h.c.\nt\\
&\ph{M_{pd} \Bigg( } +\cos\({k_y\o2}\) \psi_{d\ua}^\dag(\vk)\psi_{y\da}(\vk+\vK) -\sin\({k_y\o2}\)\psi_{y\ua}^\dag(\vk)\psi_{d\da}(\vk)+h.c.\Big) \nt\\
&+4M_{pp}\Big( \sin\({k_x\o2}\)\cos\({k_y\o2}\)\psi_{x\ua}^\dag(\vk)\psi_{y\da}(\vk) \nt\\
&\ph{+4M_{pp}\Big( }+ \cos\({k_x\o2}\)\sin\({k_y\o2}\)\psi_{y\ua}^\dag(\vk)\psi_{x\da}(\vk) + h.c.\Big)\Bigg] \,.\nt
}
The full hopping hamiltonian is now
\eq{
{\ham}_t'&=\sum_\vk{\Psi}_\vk^\dag{\hamk}(\vk) {\Psi}_\vk , &
{\Psi}_{\vk}^\dag&=\left(\psi_{d\ua}^\dag(\vk),\psi_{x\ua}^\dag(\vk),\psi_{y\ua}^\dag(\vk),\psi_{d\da}^\dag(\vk),\psi_{x\da}^\dag(\vk),\psi_{y\da}^\dag(\vk)\right)\\ 
{\hamk}(\vk)&= \begin{pmatrix}
 \hamk(\vk) & -\mathbf{M}_{\ua\da}(\vk) \\
-\mathbf{M}_{\da\ua}(\vk) & \hamk(\vk+\vK) 
 \end{pmatrix}
 &
}
where
{\renewcommand*{\arraystretch}{1.5}\eq{
\v{M}_{\ua\da}(\vk)&=\begin{pmatrix}
M_d & -2M_{pd}\cos\({k_x\o2}\) & 2M_{pd}\cos\({k_y\o2}\) \\
2M_{pd}\sin\({k_x\o2}\) & 0 & 4M_{pp}\sin\({k_x\o2}\)\cos\({k_y\o2}\) \\
-2M_{pd}\sin\({k_y\o2}\) & 4M_{pp}\cos\({k_x\o2}\)\sin\({k_y\o2}\) & 0 
\end{pmatrix}\\
\v{M}_{\da\ua}(\vk)&=\begin{pmatrix}
M_d & -2M_{pd}\sin\({k_x\o2}\) & 2M_{pd}\sin\({k_y\o2}\) \\
2M_{pd}\cos\({k_x\o2}\) & 0 & 4M_{pp}\cos\({k_x\o2}\)\sin\({k_y\o2}\) \\
-2M_{pd}\cos\({k_y\o2}\) & 4M_{pp}\sin\({k_x\o2}\)\cos\({k_y\o2}\) & 0 
\end{pmatrix}.
}}
In order to determine the parameters $M_d$, $M_{pd}$, and $M_{pp}$, we must solve the following mean-field equations:
\eq{
M_d&=\(U_d+2J\)\sum_\vk\Braket{c_{d\ua}^\dag(\vk)c_{d\da}(\vk+\vK)}\\
M_{pd}&=V_{pd}\sum_\vk\sin\({k_x\o2}\)\Braket{c_{d\ua}^\dag(\vk)c_{x\da}(\vk+\vK)}\nt\\
M_{pp}&=V_{pp}\sum_\vk\sin\({k_x\o2}\)\cos\({k_y\o2}\)\Braket{c_{x\ua}^\dag(\vk)c_{y\da}(\vk+\vK)}\nt\,.
}

As discussed in Section~\ref{sec:Tmatrix}, the presence of AF order, will mix the charge and spin channels. Naively, it follows that the number of basis functions needed will increase by a factor of four since they must now carry spin indices as well. Using the original basis functions in Table~\ref{tab:basis}, we define
\eq{\label{eqn:AFXbasis1}
\vphi_{\m\n,\s\s'}^l(\vk)=\left\{\begin{matrix*}[l]
\d_{\s\ua}\d_{\s'\ua}\,\phi^l_{\m\n}(\vk),& l=1-23\\
\d_{\s\da}\d_{\s'\da}\,\phi^{\,l-23}_{\m\n}(\vk),&l=24-46\\
\d_{\s\ua}\d_{\s'\da}\,\phi^{\,l-46}_{\m\n}(\vk),& l=47-69\\
\d_{\s\da}\d_{\s'\ua}\,\phi^{\,l-69}_{\m\n}(\vk),&l=70-92
\end{matrix*}\right. \,.
}
Alternatively, we can simply write $\vphi^l_{\m\n,\s\s'}(\vk)=\varphi^l_{ab}(\vk)$ where $a=(\m,\s)$ and $b=(\n,\s')$.
In this basis, the interaction Hamiltonian is
\eq{
\ham_{int}=&
\sum_{\vq,\vk,\vk'}\Bigg[  \sum_{l,m\in\mathcal{I}^1}\sum_{\mu\nu}\sum_{\a\b}
{ {\V}_{l}\o2} {\vphi}^{\,l}_{\m\n,\a\b}(\vk){\vphi}^{\,m}_{\m\n,\a\b}(\vk')  \psi^\dag_{\mu\a}(\vk'-\vq/2)\psi_{\mu\a}(\vk-\vq/2)\psi^\dag_{\nu\b}(\vk+\vq/2)\psi_{\nu\b}(\vk'+\vq/2)\\
	&+\sum_{l,m\in\mathcal{I}^J} {\mathcal{J}_l\o 2}\bigg( \sum_{\substack{\a\b}}\,{\vphi}^{\,l}_{dd,\a\b}(\vk){\vphi}^{\,m}_{dd,\a\b}(\vk') \big(  \psi^\dag_{d\a}(\vk'-\vq/2)\s^z_{\a\a}\psi_{d\a}(\vk-\vq/2)\psi^\dag_{d\b}(\vk+\vq/2)\s^z_{\b\b}\psi_{d\b}(\vk'+\vq/2) \nt\\
	&\ph{+\sum_{l,m\in\mathcal{I}^J} \V_l \bigg( } -2{\vphi}^{\,l}_{dd,\da\da}(\vk){\vphi}^{\,m}_{dd,\ua\ua}(\vk') \psi^\dag_{d\ua}(\vk'-\vq/2)\psi_{d\da}(\vk-\vq/2)\psi^\dag_{d\da}(\vk+\vq/2)\psi_{d\ua}(\vk'+\vq/2) \nt \\
	&\ph{+\sum_{l,m\in\mathcal{I}^J} \V_l \bigg( }- 2{\vphi}^{\,l}_{dd,\ua\ua}(\vk){\vphi}^{\,m}_{dd,\da\da}(\vk') \psi^\dag_{d\da}(\vk'-\vq/2)\psi_{d\ua}(\vk-\vq/2)\psi^\dag_{d\ua}(\vk+\vq/2)\psi_{d\da}(\vk'+\vq/2) \nt
	\bigg)\Bigg]\,.\nt
}
where $\mathcal{I}^1=\{1-19,24-42,47-65,70-88\}$ and $\mathcal{I}^J=\{20-23,43-46,66-69,89-92\}$. The parameters $\V_l$ are given by
\eq{\label{eqn:AFxintparam}
\V_l &=\left\{\begin{matrix*}[l]
\V_l',&l=1-23\\
\V_{l-23}',&l=24-46\\
\V_{l-46}',&l=47-69\\
\V_{l-69}',&l=70-92
\end{matrix*}\right.&
\V_l'&=\left\{\begin{matrix*}[l]
U_d,& l=1\\
U_p,&l=2,3\\
V_{pd},&l=4-11\\
V_{pp},&l=12-19\\
0,&\text{otherwise}
\end{matrix*}\right.\,.
}
and the Cu-Cu exchange interaction strength is simply
\eq{\label{eqn:AFxJintparam}
\mathcal{J}_l &=\left\{\begin{matrix*}[l]
\mathcal{J}_l',&l=1-23\\
\mathcal{J}_{l-23}',&l=24-46\\
\mathcal{J}_{l-46}',&l=47-69\\
\mathcal{J}_{l-69}',&l=70-92
\end{matrix*}\right.&
\mathcal{J}'_l&=\left\{
\begin{matrix*}[l]
J/4,& l=1\\
0,&\text{otherwise}
\end{matrix*}\right.\,.
}

\section{$T$-Matrix Solutions in the presence of AF order}\label{app:AFTmatrix}\label{sec:intAFX}
Here we reproduce the calculation of Section~\ref{sec:Tmatrix} in the presence of AF order. As above, the interaction vertex may be separated into an exchange and a direct part. It is given by
\eq{
{\mathbf{V}}^{\a\a',\b\b'}_{\m\m',\n\n'}(\vk,\vk';\vq)&={\mathbf{X}}^{\a\a',\b\b'}_{\m\m',\n\n'}(\vk-\vk')-{\mathbf{W}}^{\a\a',\b\b'}_{\m\m',\n\n'}(\vq)\\
&=\sum_{lm}\vphi^l_{\m\n',\a\b'}(\vk)\({X}_{lm}-{W}_{lm}(\vq)\)\vphi^m_{\m'\n,\a'\b} (\vk')\,. \nt
}
To take the different nontrivial spin behaviour into account, we will further separate both the exchange and direct vertices into a $J=0$ and a $J\neq0$ part. 

Starting with the exchange vertex, we write $\v{X}=\v{X^1}+\v{X}^J$. The $J=0$ part is given by 
\eq{
\v{X}^1=\bcX^1
\begin{pmatrix}
 \mathds{1} & 0 & 0 &0 \\
  0&\mathds{1} & 0 &0 \\
  0&0& \mathds{1} &0 \\
  0&0&0&  \mathds{1}\\
\end{pmatrix}
}
where $\mathcal{X}$ is a $23\times23$ diagonal matrix with elements $\mathcal{X}^1_{lm}=\d_{lm}\V_l'$ with $\V'_l$ given in Eq.~(\ref{eqn:AFxintparam}).
The Cu-Cu exchange term is more complicated, since it depends on the incoming and outgoing spin:
\eq{
\v{X}^J=\bcX^J
\begin{pmatrix*}[r]
 \mathds{1} & -2\mathds{1} & 0 &0\ph{-} \\
  -2\mathds{1}&\mathds{1} & 0 &0 \ph{-}\\
  0&0& -\mathds{1} &0 \ph{-}\\
  0&0&0&-\mathds{1}\ph{-}
\end{pmatrix*}
}
where $\bcX^J$ is a $23\times23$ diagonal matrix with elements $\mathcal{X}_{lm}^J=\d_{lm}\mathcal{J}_l'$ with $\mathcal{J}_l'$ given in Eq.~(\ref{eqn:AFxJintparam}).
Adding these terms, the total exchange interaction is
\eq{
\v{X}=
\begin{pmatrix}
\bb{\mathcal{X}}^1+\bcX^J & -2\bcX^J & 0 & 0 \\
-2\bcX^J & \bcX^1+\bcX^J & 0 & 0 \\
0 & 0 & \bcX^1-\bcX^J & 0 \\
0 & 0 & 0 & \bcX^1-\bcX^J 
\end{pmatrix}
}

We similarly separate the direct part into a $J=0$ and a $J\neq0$ part:
\eq{
\v{W}(\vq)=\v{W}^1(\vq)+\v{W}^J(\vq)\,.
}
The $J=0$ part is given by
\eq{
\mathbf{W}^1(\vq)=\bcW^1(\vq)\begin{pmatrix}
\id & \id & 0 & 0 \\
\id & \id & 0 &0\\
0&0&0&0\\
0&0&0&0
\end{pmatrix}
}
where $\bcW^1(\vq)$ is the same $23\times23$ matrix that was used in the case without AFM: $\mathcal{W}^1_{lm}=0$ for $l,m>3$ and for $l,m\leq3$ is given by
{\renewcommand*{\arraystretch}{1.3}\eq{\label{eqn:directint2}
\mathcal{W}^1_{lm}(\vq)=
\begin{pmatrix}
U_d & 2V_{pd}\cos(q_x/2) & 2V_{pd}\cos(q_y/2) \\
2V_{pd}\cos(q_x/2) & U_p & 4V_{pp}\cos(q_x/2)\cos(q_y/2) \\
2V_{pd}\cos(q_y/2) & 4V_{pp}\cos(q_x/2)\cos(q_y/2) & U_p
\end{pmatrix}_{lm}
\,.
}}
The Cu-Cu exchange part, $\v{W}^J(\vq)$, is given by
\eq{
\v{W}^J(\vq)=\bcW^J(\vq)
\begin{pmatrix*}[r]
\id & -\id & 0 & 0\ph{-}\\
-\id & \id & 0 & 0\ph{-}\\
0&0&-2\id&0\ph{-}\\
0&0&0&-2\id\ph{-}
\end{pmatrix*}
}
where $\bcW^J(\vq)$ is a $23\times23$ matrix with elements
\eq{
\mathcal{W}^J_{lm}(\vq)=\left\{
\begin{matrix*}[l]
{1\o2}J\(\cos q_x+\cos q_y\),& (l,m)=(1,1)\\
0,&\text{otherwise}
\end{matrix*}\right.\,.
}
The total direct interaction may thus be written as
\eq{
\v{W}(\vq)=\begin{pmatrix}
\bcW^1+\bcW^J &\bcW^1-\bcW^J & 0 & 0\\
\bcW^1-\bcW^J  & \bcW^1+\bcW^J  &0 &0\\
0&0&-2\bcW^J&0\\
0&0&0&-2\bcW^J
\end{pmatrix}
}

The Green's functions are given by diagonalizing the hopping Hamiltonian:
\eq{
\v{S}^\dag(\vk)\hamk(\vk)\v{S}(\vk)=\v{\Lambda}(\vk)
}
where ${\Lambda}_{ab}(\vk)=\d_{ab}{E}^a_{\vk}$ gives the band energies and ${\mathbf{S}}(\vk)$ is a $6\times6$ matrix of eigenvectors. The roman character indices ``$a$" indicate the pair $(\m,\s)$.
In the diagonal basis, the bare Green's function is 
\eq{
\mathcal{G}_{a}(\vk;\omega_n)={-1\o i\omega_n-(E^a(\vk)-\mu)}
}
and so the Green's function in the orbital basis is
\eq{
G_{ab}(\vk;\omega_n)=-\sum_c S^*_{ac}(\vk)S_{bc}(\vk){1\o i\omega_n- (E^c(\vk)-\m)}\,.
}

The full interaction is given by
\eq{
\Gamma_{lm}(\vq)&=V_{lm}(\vq)
+ \sum_{n,s=1}^{92}\sum_{\substack{aa'\\bb'}}\sum_{\vp,\omega_n} 
V_{ln}(\vq)\vphi^n_{ab'}(\vp)G_{bb'}(\vq-\vq/2)\vphi_{a'b}^s(\vp)G_{aa'}(\vp+\vq/2)\Gamma_{sm}(\vq)\\
&=V_{lm}(\vq)
+ \sum_{n,s=1}^{92}
V_{ln}(\vq)\Pi_{ns}(\vq)\Gamma_{sm}(\vq)\nt
}
where polarizability is defined as
\eq{
\Pi_{ns}(\vq)&=\sum_{\substack{aa'\\bb'}}\sum_{\vp,\omega_n} 
\vphi^n_{ab'}(\vp)G_{bb'}(\vq-\vq/2)\vphi_{a'b}^s(\vp)G_{aa'}(\vp+\vq/2)\\
&=
-\sum_\vp \sum_{\substack{aa'\\bb'}}\sum_{cc'} 
\vphi^n_{ab'}(\vp)\vphi_{a'b}^s (\vp)\mathcal{M}^{bb'aa'}_{c'c,\vp\vq}{f(E_{c'}(\vp-\vq/2))-f(E_c(\vp+\vq/2)) \o E_{c'}(\vp-\vq/2) - E_c(\vp+\vq/2) }\nt 
}
with
\eq{
\mathcal{M}^{bb'aa'}_{c'c,\vp\vq}=S^*_{bc'}(\vp-\vq/2)S_{b'c'}(\vp-\vq/2)S^*_{ac}(\vp+\vq/2)S_{a'c}(\vp+\vq/2)\,.
}
It follows that we seek the minimum eigenvalues and corresponding eigenvectors of
\eq{
A_{lm}(\vq)=\d_{lm}-\sum_{n,s=1}^{92}V_{lm}(\vq)\Pi_{ns}(\vq)\,.
}
\section{Symmetries}
\label{app:symmetries}

This appendix discusses the symmetries of our basis functions in the presence of antiferromagnetic order.
In particular we have to pay careful attention to the mixing of the charge density wave mode at wavevector $\vQ$
with spin density wave at wavevector $\vQ + (\pi, \pi)$.

The Hamiltonian commutes with the total $x$ spin
\eq{
S_x=\sum_\vk\sum_{\m}\(c^\dag_{\m\ua}(\vk)c_{\m\da}(\vk)+c_{\m\da}^\dag(\vk)c_{\m\ua}(\vk)\)
}
and a translation and spin inversion about the $z$-axis:
\eq{
\mathcal{A}: \quad c_{\m\ua}(\vk)\rightarrow e^{ik_{x,y}}c_{\m\ua}(\vk),\qquad c_{\m\da}(\vk)\rightarrow -e^{ik_{x,y}}c_{\m\da}(\vk) \,.
}
It follows that the Hamiltonian has the following invariant operators carrying momentum $\vq$:
\eq{
\sum_\vk &\phi^l_{\m\n}(\vk)\(c_{\m\ua}^\dag(\vk+\vq/2)c_{\n\ua}(\vk-\vq/2)+c_{\m\da}^\dag(\vk+\vq/2)c_{\n\da}(\vk-\vq/2)\)\\
\sum_\vk &\phi^l_{\m\n}(\vk)\(c_{\m\ua}^\dag(\vk+\vq/2+\vK)c_{\n\da}(\vk-\vq/2)+c_{\m\da}^\dag(\vk+\vq/2+\vK)c_{\n\ua}(\vk-\vq/2)\) \nt \,.
}
In terms of the $\psi_{\m\a}(\vk)$ operators defined above, these are written as 
\eq{
\sum_\vk&\(\phi^l_{\m\n}(\vk)\psi_{\m\ua}^\dag(\vk+\vq/2)\psi_{\n\ua}(\vk-\vq/2)+\phi^l_{\m\n}(\vk+\vK)\psi_{\m\da}^\dag(\vk+\vq/2)\psi_{\n\da}(\vk-\vq/2)\) \\
\sum_\vk&\(\phi^l_{\m\n}(\vk+\vK)e^{2i\vK\cdot\v{R}_\m}\psi_{\m\ua}^\dag(\vk+\vq/2)\psi_{\n\da}(\vk-\vq/2)+\phi^l_{\m\n}(\vk)\psi_{\m\da}^\dag(\vk+\vq/2)\psi_{\n\ua}(\vk-\vq/2)\)\nt\,.
}
Since we are only interested in the $S_x=0$ channel, we can use these symmetries to define a smaller set of basis functions than given in Eq.~\ref{eqn:AFXbasis1}.
We denote these functions $\chi_{\m\n,\a\b}^l(\vk)$ and list them in Tables \ref{tab:basisAFX2p1} and \ref{tab:basisAFX2p2}.
The invariant operator corresponding to each basis function is denoted $\bchi^l(\vq)$.
\begin{table}\begin{minipage}{\textwidth}\center
\begin{tabular}{r | c c | c }
$l$ & $\mu$ & $\nu$ &$\chi^{\,l}_{\mu\nu,\a\b}(\vk)$ \tnl\hline\hline
1 & $d$ & $d$ &	$\delta_{\mu d}\,\delta_{\nu d}\(\d_{\a\ua}\d_{\b\ua}+\d_{\a\da}\d_{\b\da}\)/\sqrt{2}$   \tnl
2 & $x$ & $x$ &	$\delta_{\mu x}\,\delta_{\nu x}\(\d_{\a\ua}\d_{\b\ua}+\d_{\a\da}\d_{\b\da}\)/\sqrt{2}$ \tnl 
3 & $y$ & $y$ &	$\delta_{\mu y}\,\delta_{\nu y}\(\d_{\a\ua}\d_{\b\ua}+\d_{\a\da}\d_{\b\da}\)/\sqrt{2}$ \tnl \hline
4 &$d$ & $x$ &  	$\delta_{\mu d}\,\delta_{\nu x}\Big(\d_{\a\ua}\d_{\b\ua}\cos\left({k_x\o2}\right)-\d_{\a\da}\d_{\b\da}\sin\left({k_x\o2}\right)\Big)$ \tnl
5 & $x$ & $d$ & 	$\delta_{\mu x}\,\delta_{\nu d}\Big(\d_{\a\ua}\d_{\b\ua}\cos\left({k_x\o2}\right)-\d_{\a\da}\d_{\b\da}\sin\left({k_x\o2}\right)\Big)$ \tnl
6 & $d$ & $x$ &	$\delta_{\mu d}\,\delta_{\nu x}\Big(\d_{\a\ua}\d_{\b\ua}\sin\left({k_x\o2}\right)+\d_{\a\da}\d_{\b\da}\cos\left({k_x\o2}\right)\Big)$ \tnl
7 & $x$ & $d$ & 	$\delta_{\mu x}\,\delta_{\nu d}\Big(\d_{\a\ua}\d_{\b\ua}\sin\left({k_x\o2}\right)+\d_{\a\da}\d_{\b\da}\cos\left({k_x\o2}\right)\Big)$\tnl\hline
8 & $d$ & $y$ &	$\delta_{\mu d}\,\delta_{\nu y}\Big(\d_{\a\ua}\d_{\b\ua}\cos\left({k_y\o2}\right)-\d_{\a\da}\d_{\b\da}\sin\left({k_y\o2}\right)\Big)$\tnl
9 & $y$ & $d$ &	$\delta_{\mu y}\,\delta_{\nu d}\Big(\d_{\a\ua}\d_{\b\ua}\cos\left({k_y\o2}\right)-\d_{\a\da}\d_{\b\da}\sin\left({k_y\o2}\right)\Big)$\tnl 
10 & $d$ & $y$ &	$\delta_{\mu d}\,\delta_{\nu y}\Big(\d_{\a\ua}\d_{\b\ua}\sin\left({k_y\o2}\right)+\d_{\a\da}\d_{\b\da}\cos\left({k_y\o2}\right)\Big)$ \tnl
11 & $y$ & $d$ &	$\delta_{\mu y}\,\delta_{\nu d}\Big(\d_{\a\ua}\d_{\b\ua}\sin\left({k_y\o2}\right)+\d_{\a\da}\d_{\b\da}\cos\left({k_y\o2}\right)\Big)$\tnl\hline
12 & $x$ & $y$ &  	$\delta_{\mu x}\,\delta_{\nu y}\,\sqrt{2}\Big(\d_{\a\ua}\d_{\b\ua}\cos\left({k_x\o2}\right)\cos\left({k_y\o2}\right)+\d_{\a\da}\d_{\b\da}\sin\left({k_x\o2}\right)\sin\left({k_y\o2}\right)\Big)$   \tnl 
13 & $y$ & $x$ &  	$\delta_{\mu y}\,\delta_{\nu x}\,\sqrt{2}\Big(\d_{\a\ua}\d_{\b\ua}\cos\left({k_x\o2}\right)\cos\left({k_y\o2}\right)+\d_{\a\da}\d_{\b\da}\sin\left({k_x\o2}\right)\sin\left({k_y\o2}\right)\Big)$   \tnl \
14 & $x$ & $y$ &  	$\delta_{\mu x}\,\delta_{\nu y}\,\sqrt{2}\Big(\d_{\a\ua}\d_{\b\ua}\sin\left({k_x\o2}\right)\sin\left({k_y\o2}\right)+\d_{\a\da}\d_{\b\da}\cos\left({k_x\o2}\right)\cos\left({k_y\o2}\right)\Big)$   \tnl
15 &  $y$ & $x$ & 	$\delta_{\mu y}\,\delta_{\nu x}\,\sqrt{2}\Big(\d_{\a\ua}\d_{\b\ua}\sin\left({k_x\o2}\right)\sin\left({k_y\o2}\right)+\d_{\a\da}\d_{\b\da}\cos\left({k_x\o2}\right)\cos\left({k_y\o2}\right)\Big)$   \tnl \
16 & $x$ & $y$ & 	$\delta_{\mu x}\,\delta_{\nu y}\,\sqrt{2}\Big(\d_{\a\ua}\d_{\b\ua}\cos\left({k_x\o2}\right)\sin\left({k_y\o2}\right)-\d_{\a\da}\d_{\b\da}\sin\left({k_x\o2}\right)\cos\left({k_y\o2}\right)\Big)$ \tnl
17 & $y$ & $x$ & 	$\delta_{\mu y}\,\delta_{\nu x}\,\sqrt{2}\Big(\d_{\a\ua}\d_{\b\ua}\cos\left({k_x\o2}\right)\sin\left({k_y\o2}\right)-\d_{\a\da}\d_{\b\da}\sin\left({k_x\o2}\right)\cos\left({k_y\o2}\right)\Big)$  \tnl\
18 & $x$ & $y$ & 	$\delta_{\mu x}\,\delta_{\nu y}\,\sqrt{2}\Big(\d_{\a\ua}\d_{\b\ua}\sin\left({k_x\o2}\right)\cos\left({k_y\o2}\right)-\d_{\a\da}\d_{\b\da}\cos\left({k_x\o2}\right)\sin\left({k_y\o2}\right)\Big)$ \tnl
19 & $y$ & $x$ & 	$\delta_{\mu y}\,\delta_{\nu x}\,\sqrt{2}\Big(\d_{\a\ua}\d_{\b\ua}\sin\left({k_x\o2}\right)\cos\left({k_y\o2}\right)-\d_{\a\da}\d_{\b\da}\cos\left({k_x\o2}\right)\sin\left({k_y\o2}\right)\Big)$ \tnl \hline
20 & $d$ & $d$ & 	$\delta_{\mu d}\,\delta_{\nu d}\(\d_{\a\ua}\d_{\b\ua}-\d_{\a\da}\d_{\b\da}\)\left(\cos{k_x}-\cos k_y \right)/\sqrt{2}$ \tnl
21 & $d$ & $d$ & 	$\delta_{\mu d}\,\delta_{\nu d}\(\d_{\a\ua}\d_{\b\ua}-\d_{\a\da}\d_{\b\da}\)\left(\cos{k_x}+\cos k_y \right)/\sqrt{2}$  \tnl
22 & $d$ & $d$ & 	$\delta_{\mu d}\,\delta_{\nu d}\(\d_{\a\ua}\d_{\b\ua}-\d_{\a\da}\d_{\b\da}\)\left(\sin{k_x}-\sin k_y \right)/\sqrt{2}$  \tnl
23 & $d$ & $d$ & 	$\delta_{\mu d}\,\delta_{\nu d}\(\d_{\a\ua}\d_{\b\ua}-\d_{\a\da}\d_{\b\da}\)\left(\sin{k_x}+\sin k_y \right)/\sqrt{2}$ 
\end{tabular}
\end{minipage}
\caption{The first 23 of 46 basis functions for the $S_x=0$ channel for the case of an AFM in the $x$-direction. For each $l$-index,  $\chi^l_{\mu\nu,\a\b}(\vk)$ is nonzero only for the $\mu\nu-$pair given in the second and third columns of each table. The full function is shown in the fourth column.}\label{tab:basisAFX2p1}
\end{table}
\begin{table}
\begin{minipage}{\textwidth}\center
\begin{tabular}{r | c c | c }
$l$ & $\mu$ & $\nu$ &$\chi^{\,l}_{\mu\nu,\a\b}(\vk)$ \tnl\hline\hline
24 & $d$ & $d$ &	$\delta_{\mu d}\,\delta_{\nu d}\(\d_{\a\ua}\d_{\b\da}+\d_{\a\da}\d_{\b\ua}\)/\sqrt{2}$   \tnl
25 & $x$ & $x$ &	$\delta_{\mu x}\,\delta_{\nu x}\(\d_{\a\ua}\d_{\b\da}-\d_{\a\da}\d_{\b\ua}\)/\sqrt{2}$ \tnl 
26 & $y$ & $y$ &	$\delta_{\mu y}\,\delta_{\nu y}\(\d_{\a\ua}\d_{\b\da}-\d_{\a\da}\d_{\b\ua}\)/\sqrt{2}$ \tnl \hline
27 &$d$ & $x$ &  	$\delta_{\mu d}\,\delta_{\nu x}\Big(\d_{\a\ua}\d_{\b\da}\cos\left({k_x\o2}\right)+\d_{\a\da}\d_{\b\ua}\sin\left({k_x\o2}\right)\Big)$ \tnl
28& $x$ & $d$ & 	$\delta_{\mu x}\,\delta_{\nu d}\Big(\d_{\a\ua}\d_{\b\da}\cos\left({k_x\o2}\right)-\d_{\a\da}\d_{\b\ua}\sin\left({k_x\o2}\right)\Big)$ \tnl
29 & $d$ & $x$ &	$\delta_{\mu d}\,\delta_{\nu x}\Big(\d_{\a\ua}\d_{\b\da}\sin\left({k_x\o2}\right)-\d_{\a\da}\d_{\b\ua}\cos\left({k_x\o2}\right)\Big)$ \tnl
30 & $x$ & $d$ & 	$\delta_{\mu x}\,\delta_{\nu d}\Big(\d_{\a\ua}\d_{\b\da}\sin\left({k_x\o2}\right)+\d_{\a\da}\d_{\b\ua}\cos\left({k_x\o2}\right)\Big)$\tnl\hline
31 & $d$ & $y$ &	$\delta_{\mu d}\,\delta_{\nu y}\Big(\d_{\a\ua}\d_{\b\da}\cos\left({k_y\o2}\right)+\d_{\a\da}\d_{\b\ua}\sin\left({k_y\o2}\right)\Big)$\tnl
32 & $y$ & $d$ &	$\delta_{\mu y}\,\delta_{\nu d}\Big(\d_{\a\ua}\d_{\b\da}\cos\left({k_y\o2}\right)-\d_{\a\da}\d_{\b\ua}\sin\left({k_y\o2}\right)\Big)$\tnl 
33 & $d$ & $y$ &	$\delta_{\mu d}\,\delta_{\nu y}\Big(\d_{\a\ua}\d_{\b\da}\sin\left({k_y\o2}\right)-\d_{\a\da}\d_{\b\ua}\cos\left({k_y\o2}\right)\Big)$ \tnl
34 & $y$ & $d$ &	$\delta_{\mu y}\,\delta_{\nu d}\Big(\d_{\a\ua}\d_{\b\da}\sin\left({k_y\o2}\right)+\d_{\a\da}\d_{\b\ua}\cos\left({k_y\o2}\right)\Big)$\tnl\hline
35 & $x$ & $y$ &  	$\delta_{\mu x}\,\delta_{\nu y}\,\sqrt{2}\Big(\d_{\a\ua}\d_{\b\da}\cos\left({k_x\o2}\right)\cos\left({k_y\o2}\right)-\d_{\a\da}\d_{\b\ua}\sin\left({k_x\o2}\right)\sin\left({k_y\o2}\right)\Big)$   \tnl 
36 & $y$ & $x$ &  	$\delta_{\mu y}\,\delta_{\nu x}\,\sqrt{2}\Big(\d_{\a\ua}\d_{\b\da}\cos\left({k_x\o2}\right)\cos\left({k_y\o2}\right)-\d_{\a\da}\d_{\b\ua}\sin\left({k_x\o2}\right)\sin\left({k_y\o2}\right)\Big)$   \tnl 
37 & $x$ & $y$ &  	$\delta_{\mu x}\,\delta_{\nu y}\,\sqrt{2}\Big(\d_{\a\ua}\d_{\b\da}\sin\left({k_x\o2}\right)\sin\left({k_y\o2}\right)-\d_{\a\da}\d_{\b\ua}\cos\left({k_x\o2}\right)\cos\left({k_y\o2}\right)\Big)$   \tnl
38 &  $y$ & $x$ & 	$\delta_{\mu y}\,\delta_{\nu x}\,\sqrt{2}\Big(\d_{\a\ua}\d_{\b\da}\sin\left({k_x\o2}\right)\sin\left({k_y\o2}\right)-\d_{\a\da}\d_{\b\ua}\cos\left({k_x\o2}\right)\cos\left({k_y\o2}\right)\Big)$   \tnl 
39 & $x$ & $y$ & 	$\delta_{\mu x}\,\delta_{\nu y}\,\sqrt{2}\Big(\d_{\a\ua}\d_{\b\da}\cos\left({k_x\o2}\right)\sin\left({k_y\o2}\right)+\d_{\a\da}\d_{\b\ua}\sin\left({k_x\o2}\right)\cos\left({k_y\o2}\right)\Big)$ \tnl
40 & $y$ & $x$ & 	$\delta_{\mu y}\,\delta_{\nu x}\,\sqrt{2}\Big(\d_{\a\ua}\d_{\b\da}\cos\left({k_x\o2}\right)\sin\left({k_y\o2}\right)+\d_{\a\da}\d_{\b\ua}\sin\left({k_x\o2}\right)\cos\left({k_y\o2}\right)\Big)$  \tnl
41 & $x$ & $y$ & 	$\delta_{\mu x}\,\delta_{\nu y}\,\sqrt{2}\Big(\d_{\a\ua}\d_{\b\da}\sin\left({k_x\o2}\right)\cos\left({k_y\o2}\right)+\d_{\a\da}\d_{\b\ua}\cos\left({k_x\o2}\right)\sin\left({k_y\o2}\right)\Big)$ \tnl
42 & $y$ & $x$ & 	$\delta_{\mu y}\,\delta_{\nu x}\,\sqrt{2}\Big(\d_{\a\ua}\d_{\b\da}\sin\left({k_x\o2}\right)\cos\left({k_y\o2}\right)+\d_{\a\da}\d_{\b\ua}\cos\left({k_x\o2}\right)\sin\left({k_y\o2}\right)\Big)$ \tnl \hline
43 & $d$ & $d$ & 	$\delta_{\mu d}\,\delta_{\nu d}\(\d_{\a\ua}\d_{\b\da}-\d_{\a\da}\d_{\b\ua}\)\left(\cos{k_x}-\cos k_y \right)/\sqrt{2}$ \tnl
44 & $d$ & $d$ & 	$\delta_{\mu d}\,\delta_{\nu d}\(\d_{\a\ua}\d_{\b\da}-\d_{\a\da}\d_{\b\ua}\)\left(\cos{k_x}+\cos k_y \right)/\sqrt{2}$  \tnl
 45& $d$ & $d$ & 	$\delta_{\mu d}\,\delta_{\nu d}\(\d_{\a\ua}\d_{\b\da}-\d_{\a\da}\d_{\b\ua}\)\left(\sin{k_x}-\sin k_y \right)/\sqrt{2}$  \tnl
46 & $d$ & $d$ & 	$\delta_{\mu d}\,\delta_{\nu d}\(\d_{\a\ua}\d_{\b\da}-\d_{\a\da}\d_{\b\ua}\)\left(\sin{k_x}+\sin k_y \right)/\sqrt{2}$ 
\end{tabular}
\end{minipage}
\caption{The second 23 of 46 basis functions for the $S_x=0$ channel for the case of an AFM in the $x$-direction. For each $l$-index,  $\chi^l_{\mu\nu,\a\b}(\vk)$ is nonzero only for the $\mu\nu-$pair given in the second and third columns of each table. The full function is shown in the fourth column.}\label{tab:basisAFX2p2}
\end{table}

We can get to the new basis from the $\vphi^l_{\m\n,\a\b}(\vk)$ by an appropriate projection matrix $\v{B}$. Given the following definitions
\eq{
\v{D}^s_{pd}&=\begin{pmatrix}
0&-\id_{2\times2}\\\id_{2\times2}& 0 \end{pmatrix}
&
\v{D}_{pd}^t&=\begin{pmatrix}
0&\s^z\\-\s^z&0\end{pmatrix}\\
\v{D}_{pp}&=\begin{pmatrix}
0&0&0&\id_{2\times2}\\
\id_{2\times2}&0&0&0\\
0&0&-\id_{2\times2}&0\\
0&-\id_{2\times2}&0&0
\end{pmatrix}\nt\\
\v{D}_{s}&=\begin{pmatrix}
\id_{3\times3}&0&0&0&0\\
0&\v{D}^s_{pd}&0&0&0\\
0&0&\v{D}^s_{pd}&0&0\\
0&0&0&\v{D}_{pp}&0\\
0&0&0&0&-\id_{4\times4}
\end{pmatrix}
&
\v{D}_{t}&=\begin{pmatrix}
1&0&0&0&0&0\\
0&-\id_{2\times2}&0&0&0&0\\
0&0&\v{D}^t_{pd}&0&0&0\\
0&0&0&\v{D}^t_{pd}&0&0\\
0&0&0&0&-\v{D}_{pp}&0\\
0&0&0&0&0&-\id_{4\times4}
\end{pmatrix}\nt \,.
}
we can define $\v{B}$ in terms of a rotation matrix $\v{U}$ and projector matrix $\v{P}$ :
\eq{
\v{U}&= \begin{pmatrix}
\id_{23\times23} &0&0&0\\
0&\v{D}_s &0&0 \\
0&0&\id_{23\times23}&0\\
0&0&0&\v{D}_t \end{pmatrix} 
&
\v{P}& = {1\o \sqrt{2}} \begin{pmatrix}
\id_{23\times23} &\id_{23\times23} &0 &0 \\
0&0&\id_{23\times23} &\id_{23\times23} \end{pmatrix}
}
\eq{
\v{B}=\v{P}\v{U} \,.
}
That is, (dropping the orbital and spin indices), we have
\eq{
\chi^l(\vk)=\sum_{m=1}^{92}B_{lm}\vphi^m(\vk)\,.
}
In the new basis, the interaction vertices defined in Appendix~\ref{sec:intAFX} must be rewritten as
\eq{
\mathbf{\bar{V}}(\vq) = \v{B} \v{V}(\vq)\v{B}^T \,.
}
The remainder of the calculation presented in Appendix~\ref{sec:intAFX} is identical save with $46\times46$ matrices instead of $92\times92$.

The Hamiltonian is additionally invariant under the transformation 
\eq{\label{eqn:AFparity}
\mathcal{B}:\quad c_{\m\a}(\vk)\rightarrow\eta_{\m\n}c_{\n\a}(-\vk), \qquad\quad \eta_{\m\n}=\text{diag}(1,-1,-1)\,.
}
(The $\eta_{\m\n}$ factor is due to the gauge choice of Eq.~(\ref{eqn:fourdef})). Combined with complex conjugation, there are 24 invariant operators remaining, which are listed in Table \ref{tab:AFparity}.
Instead of working directly with these operators, we instead work with those given in Tables \ref{tab:basisAFX2p1} and ~\ref{tab:basisAFX2p2} and afterwards ensure that all all eigenvectors satisfy this symmetry.
\begin{table}
\begin{minipage}{\textwidth}\center
\begin{tabular}{c|c}
\multicolumn{2}{c}{Parity invariant operators}\\\hline
Singlet & Triplet \\ \hline\hline\\ [-2.ex]
$\bchi^1$ 	& $\bchi^{24}$	\tnl
$\bchi^2$ &  \tnl
$\bchi^3$ &  \tnl\hline\\ [-2.ex]
${1\o\sqrt{2}}\(\bchi^4-\bchi^5\)$	&${1\o\sqrt{2}}\(\bchi^{27}+\bchi^{30}\)$ \tnl
${1\o\sqrt{2}}\(\bchi^6+\bchi^7\)$ 	&${1\o\sqrt{2}}\(\bchi^{28}+\bchi^{29}\)$ \tnl\hline\\ [-2.ex]
${1\o\sqrt{2}}\(\bchi^8-\bchi^9\)$	& ${1\o\sqrt{2}}\(\bchi^{31}+\bchi^{34}\)$\tnl
${1\o\sqrt{2}}\(\bchi^{10}+\bchi^{11}\)$	& ${1\o\sqrt{2}}\(\bchi^{32}+\bchi^{33}\)$\tnl\hline\\ [-2.ex]
${1\o\sqrt{2}}\(\bchi^{12}+\bchi^{13}\)$	&${1\o\sqrt{2}}\(\bchi^{35}+\bchi^{38}\)$ \tnl
${1\o\sqrt{2}}\(\bchi^{14}+\bchi^{15}\)	$	&${1\o\sqrt{2}}\(\bchi^{36}+\bchi^{37}\)$ \tnl
${1\o\sqrt{2}}\(\bchi^{16}-\bchi^{17}\)$	&${1\o\sqrt{2}}\(\bchi^{39}+\bchi^{42}\)$\tnl
${1\o\sqrt{2}}\(\bchi^{18}-\bchi^{19}\)$	&${1\o\sqrt{2}}\(\bchi^{40}+\bchi^{41}\)$ \tnl\hline\\ [-2.ex]
$\bchi^{20}$ & $\bchi^{45}$ \tnl
$\bchi^{21} $ & $\bchi^{46}$
\end{tabular}
\end{minipage}
\caption{Operators invariant under the action of $\mathcal{B}\mathcal{K}$, where $\mathcal{B}$ is defined in Eq.~(\ref{eqn:AFparity}) and $\mathcal{K}$ is complex conjugation. (The $\vq$ dependence of the $\bchi^l$'s is suppressed).}\label{tab:AFparity}
\end{table}

\bibliographystyle{apsrev4-1}
\bibliography{draft1}

\printfigures

\end{document}